\documentclass[fleqn,usenatbib]{mnras}

\usepackage[utf8]{inputenc}
\usepackage{ae,aecompl}

\usepackage{mathptmx}

\usepackage{amsmath}
\usepackage{amssymb}
\usepackage{graphicx}
\usepackage[usenames,dvipsnames]{color}

\usepackage{xspace}

\usepackage{enumitem}
\usepackage{enumerate}
\setlist{listparindent=\parindent,leftmargin=*}

\definecolor{webgreen}{rgb}{0,.5,0}
\newcommand{\green}[1]{#1}

\DeclareMathOperator*{\dv}{d \!}

\newcommand{\glf}{\textsc{galform}\xspace}

\newcommand{\mc}[3]{\multicolumn{#1}{#2}{#3}}

\newcommand{\betadisk}{\beta_{0,\text{disc}}}
\newcommand{\betaburst}{\beta_{0,\text{burst}}}
\newcommand{\alphahot}{\alpha_\text{hot}}
\newcommand{\alphacool}{\alpha_\text{cool}}
\newcommand{\fstab}{f_{\rm stab}}
\newcommand{\nusf}{\nu_{0,\text{sf}}}
\newcommand{\alphareheat}{\alpha_\text{reheat}}
\newcommand{\msun}{\,\text{M}_\odot}

\hypersetup{
pdftitle={Constraints on galaxy formation models from the galaxy stellar mass
  function and its evolution},
pdfauthor={
            L.~F.~S.~Rodrigues (orcid.org/0000-0002-3860-0525),
            I.~R.~Vernon (orcid.org/0000-0002-9161-9946),
            R.~G.~Bower (orcid.org/0000-0002-5215-6010)
          }
}

\title[Constraining galaxy formation using the GSMF]
{Constraints on galaxy formation models from the galaxy stellar mass
  function and its evolution}
\author[L.~F.~S.~Rodrigues, I.~Vernon, R.~G.~Bower]
{
Luiz~Felippe~S.~Rodrigues$^{1,2}$\thanks{Email: luiz.rodrigues@newcastle.ac.uk},
Ian~Vernon$^3$ and Richard~G.~Bower$^4$
\\
$^{1}$School of Mathematics and Statistics, University of Newcastle, Newcastle
upon Tyne, NE1 7RU, UK\\
$^{2}$Centro Universitário Belas Artes de São Paulo, Rua Dr Álvaro Alvim, 90,
São Paulo, SP, 04018-010, Brazil\\
$^{3}$Department of Mathematical Sciences, University
of Durham, South Road, Durham, DH1 3LE, UK\\
$^{4}$Institute for Computational Cosmology, Department of Physics, University
of Durham, South Road, Durham, DH1 3LE, UK
}

\pagerange{\pageref{firstpage}--\pageref{lastpage}}
\pubyear{2016}
\date{Accepted for publication in MNRAS}

\begin{document}
\label{firstpage}

\maketitle

\begin{abstract}
We explore the parameter space of the semi-analytic galaxy formation model \glf,
studying the constraints imposed by measurements of the galaxy stellar mass function
(GSMF) and its evolution. We use the Bayesian Emulator method to quickly eliminate
vast implausible volumes of the parameter space and zoom in on the most interesting
regions, allowing us to identify a set of models that match the observational data
within
model uncertainties.
We find that the GSMF strongly constrains parameters related to
quiescent star formation in discs, stellar and AGN feedback and
threshold for disc instabilities, but
weakly restricts other parameters. Constraining the model
using
local data alone does not usually select models that match the evolution of
the
GSMF
well. Nevertheless, we show that a small subset of models provides
acceptable match to GSMF data out to redshift 1.5.
We explore the physical
significance of the parameters of these models, in particular exploring whether the
model
provides a
better description if the mass loading of the galactic winds
generated by starbursts ($\beta_{0,\text{burst}}$) and quiescent disks
($\beta_{0,\text{disc}}$) is different. Performing a principal component analysis of
the plausible volume of the parameter space, we write a set of relations between
parameters obeyed by plausible models with respect to
GSMF evolution. We find
that while $\beta_{0,\text{disc}}$ is strongly constrained by GSMF evolution data,
constraints on $\beta_{0,\text{burst}}$ are weak.
\green{Although it is possible to find plausible models for which
$\beta_{0,\text{burst}} = \beta_{0,\text{disc}}$,
most plausible models have $\beta_{0,\text{burst}}>\beta_{0,\text{disc}}$, implying
-- for these -- larger SN feedback efficiency at higher redshifts.}
\end{abstract}

\section{Introduction}

Semi-analytic models of galaxy formation (SAMs) are well established tools for
exploring galaxy formation scenarios in their cosmological context. The problem of
how galaxies form and evolve is described by a set of coupled differential equations
dealing with well-defined astrophysical processes. These are driven by dark
matter halo merger trees that determine the source terms in the equation network
\citep[for reviews see e.g.][]{BensonReview,Baugh2006,SomervilleDave2015}.
Due to the approximate nature of the methods used in these simulations, and the
uncertainties in the physical process that are modelled, these models include a large
number of uncertain parameters. While order of magnitude estimates for these
parameters can be made, their precise values must be determined by comparison
to observational data.

Traditionally, parameter values have been set through a trial-and-error approach,
where the galaxy formation modeller varies an individual parameter developing
intuition about its effects on the model predictions for a particular observable and
then uses this understanding to select a parameter set that gives a good description
of the observations.
Despite its simplicity, and obvious limitations, this procedure has led to
substantial progress in the field.
Recently, however, several papers have employed more rigorous statistical methods to
explore the
high dimensional parameter space systematically
\citep{Kampakoglou2008,Henriques2009,Bower2010,Henriques2013,Benson2014,Lu2014,
Henriques2015}. Such approaches
provide a richer analysis, and seek to identify the regions of
parameter space that are in agreement with observational data, and not just to find
optimal parameter values. This therefore informs as to the uniqueness of the
parameter choices, and provides understanding of the degeneracies between different
parameters.

In this work we study which constraints are imposed on the semi-analytic model
\glf by the observations of galaxy stellar mass function (GSMF).  We first consider
the constraints imposed by local observations and then investigate how the
parameters are further constrained by the introduction of high
redshift data. This makes powerful
use of  the iterative emulator technique described by \citet{Bower2010}, which
provides an efficient way of probing a high dimensional parameter space.
Importantly, the method allows additional constraints to be added in post-processing.
Thus, we start by finding the region in the parameter space which contains models
that produce a good match to the local Universe GSMF.
This region is, then, further probed to check whether a match to higher
redshift data is possible.
By analysing 2D projections of the plausible models sub-volume and performing a
principal component analysis of it, we are able to study the degeneracies and
interactions between the most constrained parameters. We note that typical
approaches to analysing comparable models using Bayesian MCMC require millions of
model runs (at least), while the approach used here, which utilises Bayesian
emulation, only required tens of thousands of model runs, representing a substantial
improvement in efficiency.

Closely reproducing the observed high-redshift galaxy mass function
\citep{Cirasuolo2010, Henriques2013} is problematic for many galaxy formation models.
\citet{Henriques2013}, for example, concludes that the effectiveness of galaxy
feedback (specifically the re-incorporation time of expelled gas) must depend on
\green{the virial mass of the dark matter halo
on the basis of a Monte Carlo exploration of the parameter space of their model.
This is not fully satisfactory, however, since one would expect the re-incorporation
time to be physically related to the halo dynamical time and not the halo mass.
}

In this paper, we explore an appealing and well-motivated alternative. Observations
of galaxy winds \citep[eg.,][]{heckman1990, martin2012} suggest that the effective
mass loading is strongly dependent on the surface density of star formation. It
appears that efficient outflows are more readily generated when star formation
occurs in dense bursts than when the star formation occurs in a smooth
and quiescent disk.  These observations motivate a more careful exploration of the
treatment of galaxy winds from starburst and quiescent disks, and in this paper we
parametrize the mass loading of the wind independently in these two cases. This may
naturally resolve the difficulty presented by observations of the high redshift GSMF
since the cosmic star formation rate density may be more dominated by starbursts at
high redshift, while it is dominated by quiescent star formation at low redshift
\citep{Malbon2007}.

This paper is organized as follows.
In \S\ref{sec:model} we describe the galaxy formation model, specifying
which parameters were varied and briefly reviewing the physical meaning of the
most relevant of them.
In \S\ref{sec:emulator} the iterative history matching methodology is
reviewed.
In \S\ref{sec:z0} we present our results for the matching to the local
GSMF.
In \S\ref{sec:high_z} we examine the effects of including higher redshift
data.
In \S\ref{sec:subspace}, 2D projections of the parameter space are
analysed.
In \S\ref{sec:pca}, the results of a principal component analysis of the
non-implausible volume are shown.
Finally, in \S\ref{sec:summary}, we summarize our conclusions.

\section{Galaxy formation model}
\label{sec:model}

The basis of this paper is the semi-analytic model \glf, first introduced by
\citet{Cole2000}. Our starting point is the model discussed by \citet[][hereafter
GP14]{Gonzalez-Perez2014}, which re-calibrates the version described in
\citet{Lagos2012}
 to match observational data taking into account the
best-fitting cosmological parameters obtained by WMAP7 \citep{WMAP7}.
The model of \citet{Lagos2012} is itself a development of the version presented
by \citet{Bower2006} -- which introduced AGN feedback and disc instabilities to
the original \glf model -- introducing a modified prescription for star formation
in galaxy discs (\S\ref{sec:BR_starformation}, see \citealt{Lagos2011sfl} for an
in-depth discussion).

We note that there is now a more modern variant of the \glf model which differs from
the base model used here. This model, described {comprehensively}
by \citet{Lacey2016}, assumes two initial mass functions (IMFs), one for quiescent
star formation and a different one for star bursts
{-- an approach which improves the model predictions for number counts and
redshift distribution of sub-millimetre galaxies.
The model presented here assumes a universal IMF, which considerably simplifies
comparison with the galaxy stellar mass function. A universal IMF is compatible with
direct observational measurements: see \citet{Bastian2010} and \citet{Smith2015} for
a recent discussion.}

\subsection{Differences from GP14}

Although the model we use here is based on GP14, there are a number of
small, but important, differences.
Firstly, the merger trees in the present study were constructed using the Monte Carlo
algorithm described by \citet{Parkinson2008}, which is based on the
Extended Press-Schechter theory \citep{Bower1991,LaceyCole1993}, while
GP14 uses merger trees extracted from a Millenium-class N-body
simulation \citep{Guo2013}.
\green{-- t}he use of Monte Carlo merger trees allows \glf to run significantly faster since
it is possible to control the number of haloes with a given final mass in the
simulation, whereas in the case of the N-body trees, most of the computational time
is spent on over-represented small mass haloes.
In GP14, ram-pressure stripping is modelled by
completely and instantaneously removing the hot gas halo when a galaxy becomes a
satellite. Here we follow the same prescription as \citet{Font2008}, which uses the
\citet{McCarthy2008} ram-pressure stripping model that is based on hydrodynamic
simulations \green{-- a} similar update to the model is used in \citet{Lagos2014}.
\green{
Finally, the present model adopts the IMF obtained by \citet{ChabrierIMF}, while GP14
uses a \citet{KennicuttIMF} IMF.
}
\subsection{Varied parameters}
\label{sec:parameters}

\begin{table*}
\begin{center}
\caption{
Parameters varied in this work, the physical processes  and their ranges.
For reference, values of these parameters used in GP14 are shown.
}
\begin{tabular}{cclccccc}
\hline
Process modelled& Section &Parameter name [units]&\mc{2}{c}{Range}  &
GP14  & Scaling\\
\hline
\hline
Star formation (quiescent)  & \S\ref{sec:BR_starformation} & $\nu_\text{sf}$ [Gyr$^{-1}$] & 0.025 & 1.0 & 0.5 & lin\\
                            &                             &$P_\text{sf}/k_\text{B}$ [cm$^{-3}$K] & $1\times10^4$& $5\times10^4$ & $1.7\times10^4$ & log\\
                            &                             &$\beta_\text{sf}$ & 0.65 & 1.10 & 0.8 & lin\\
 \hline
Star formation (bursts)     & \S\ref{sec:burst_starformation}
&$f_\text{dyn}$ & 1.0 & 100.0 & 10 & log\\
                            &                  &$\tau_\text{min,burst}$ [Gyr]  & $10^{-3}$ & 1 & 0.05 & log\\
\hline
SNe feedback & \S\ref{sec:SN_feedback}&$\alpha_\text{hot}$ & 1.0 & 3.7 & 3.2 & lin\\
             &                        & $\betaburst$ & 0.5 & 40.0 & 11.16 & lin\\
             &                        &$\betadisk$ & 0.5 & 40.0 & 11.16 & lin\\
             &                        &$\alpha_\text{reheat}$ & 0.15 & 1.5 & 1.26027 & lin\\
\hline
AGN feedback & \S\ref{sec:AGN_feedback} &$\alpha_\text{cool}$ & 0.1 & 2.0 & 0.6 & log\\
             &                         &$\epsilon_\text{edd}$ & 0.004 & 0.1 & 0.03979 & log\\
             &                         &$f_\text{smbh}$ & 0.001 & 0.01 & 0.005 & lin\\ \hline

Galaxy mergers &
&$f_\text{burst}$ & 0.01 & 0.5 & 0.1  & log\\
               &                   &$f_\text{ellip}$ & 0.01 & 0.5 & 0.3 & log\\
\hline
Disk stability & \S\ref{sec:stability}&$f_\text{stab}$ & 0.61 & 1.1 & 0.8 & lin\\
\hline
Reionization &
& $V_\text{cut}$ [km$\,$s$^{-1}$] & 20 & 60 & 30 & lin\\
             &                         & $z_\text{cut}$ & 5 & 15 & 10 & lin \\
\hline
Metal enrichment &                    &$p_\text{yield}$ & 0.02 & 0.05 & 0.021 & lin\\
\hline
Ram pressure stripping &
& $\epsilon_\text{strip}$ & 0.01 & 0.99 & n/a & lin\\
                       &                        & $\alpha_\text{rp}$      & 1.0  &
3.0 & n/a  & lin\\
\hline
\end{tabular}

\label{tab:parameters}
\end{center}
\end{table*}

The semi-analytic approach to the problem of galaxy formation relies on a large
number of parameters which codify the
uncertainties associated with the many astrophysical processes involved.
Since the emulator technique allows us to survey a parameter space of high
dimensionality both quickly and at a relatively low computational cost, we are
able to vary parameters simultaneously. One should bear in mind
that varying a larger number of parameters in the present approach corresponds
to a more \emph{conservative} choice, since it requires less \textit{a priori}
assumptions about the role of each parameter.

We varied 20 parameters, all of which are listed, together with their ranges,
in table \ref{tab:parameters}. We outline the physical meaning of
parameters related to star formation and feedback in the subsections
below, for further details, we refer the reader to the original
papers, and to \citet{Lacey2016}.

For the purposes of sampling, computations of volumes and principal
components analysis, the parameters were rescaled to $[-1,1]    $ within the
initial range, either linearly,
\begin{equation}
\label{eq:scale_lin}
 p^{(s)} = 2\left(\frac{p-p_\text{min}}{p_\text{max}-p_\text{min}}\right)-1\,,
\end{equation}
or logarithmically,
\begin{equation}
\label{eq:scale_log}
 p^{(l)} =
2\left[\frac{\log_{10}(p/p_\text{min})}{\log_{10}(p_\text{max}/p_\text{min})}
\right]-1\,.
\end{equation}
The scaling used is also listed in table \ref{tab:parameters}.

\subsubsection{Quiescent star formation}

It is assumed in the model that the surface density of the star formation rate is set by the surface density of molecular gas (see \citealt{Lagos2011sfl} and references
therein),
\label{sec:BR_starformation}
\begin{equation}
\dot \Sigma_\star = \nusf \,\Sigma_\text{mol} =\nusf
f_\text{mol} \Sigma_\text{gas}\label{eq:BR_starformation}\,,
\end{equation}
where \green{$\Sigma_\text{gas}$ is the surface density of cold gas in the disc and}
the fraction of molecular hydrogen,
$f_\text{mol}=R_\text{mol}/(R_\text{mol}+1)$, is computed using the pressure
relation of \citet{BlitzRosolowsky2006}
\begin{equation}
 R_\text{mol} = \left(\frac{P_\text{ext}}{P_\text{sf}}\right)^{\beta_\text{sf}}
 \label{eq:BR_Rmol}\,,
\end{equation}
with
\begin{equation}
 P_\text{ext}=\frac{\upi}{2} G \Sigma_\text{gas}\left[ \Sigma_\text{gas} +\left(
\frac{\sigma_\text{gas}}{\sigma_\star}\right)\Sigma_\star \right]\,.
\end{equation}

\subsubsection{Star formation bursts}

During a starburst the star formation rate is set to
\label{sec:burst_starformation}
\begin{equation}
 \text{SFR}_\text{burst} = \frac{M_\text{gas,bulge}}{\tau_{\star, \text{burst}}}
\end{equation}
with
\begin{equation}
 \tau_{\star, \text{burst}} = \max\left( f_\text{dyn}\tau_\text{dyn},\,\,
            \tau_{\text{min,burst}}  \right)\label{eq:burst}
\end{equation}
where $\tau_\text{dyn}$ is the dynamical time of the newly formed spheroid and
$f_\text{dyn}$ and $\tau_{\text{min,burst}}$ are model parameters.

\subsubsection{Supernovae feedback}
\label{sec:SN_feedback}
The outflow of gas from the disc or the bulge of a galaxy is modelled
using
\begin{equation}\label{eq:alphahot}
\dot M_\text{out,disc/burst} = \beta \times \text{SFR}_\text{disc/burst}
\end{equation}
where $\text{SFR}_\text{disc/burst}$ are the total star formation rates in the
quiescent and starburst cases and $\beta$ is the \emph{mass loading}, given by
\begin{equation}
\beta = \beta_{0,\text{disc/burst}}
\left(\frac{V_\text{disc/bulge}}{200\,\text{km}\,\text{s}^{-1}}
\right)^{-\alphahot}
\label{eq:beta}
\end{equation}
where $V_\text{disc/bulge}$ are the circular velocity associated with
with the disc (in the quiescent case) or with the newly formed spheroid
(bulge) component (in a starburst).

In previous \glf works the mass loadings associated with discs and bursts were
assumed to share the same normalization, i.e. $\betaburst=\betadisk = \beta_0$.
This assumption was relaxed in the present work. The notation in previous works was
also slightly different: the equivalent parameter
\begin{equation}
V_\text{hot}~\equiv~(200\,\text{km}\,\text{s}^{-1})~\times~\beta_0^{
-1/\alphahot}
\end{equation}
was used instead.

The outflowed gas is assumed to be once more available to cool and form stars
on a time-scale
\begin{equation}
 t_\text{reinc} = \frac{\tau_\text{halo}}{\alphareheat},
\label{eq:reheat}
\end{equation}
where $\tau_\text{halo}$ is the dynamical time of the halo.
\green{The amount of cold gas available (as well as the amount of stars formed) is
determined by simultaneously solving for both the star formation rate and the
outflow rate.}

\subsubsection{AGN feedback}
\label{sec:AGN_feedback}

The model assumes the cooling of gas from the hot gas halo can be disrupted by
the injection of energy by the AGN. This is assumed to happen only at haloes
under `quasi-hydrostatic equilibrium', defined by
\begin{equation}
 t_\text{cool}(r_\text{cool}) > \alphacool^{-1} t_\text{ff}(r_\text{cool})
\label{eq:tcool}
\end{equation}
where $t_\text{cool}$ and $r_\text{cool}$ are the cooling time and radius and
$t_\text{ff}$ is the free fall time. Thus, the parameter $\alphacool$ determines
the halo mass at which AGN feedback is effective (i.e. lower values of
$\alphacool$ implies AGN feedback active in smaller mass haloes).

The cooling of gas from the hot gas halo is interrupted if a galaxy satisfies
equation \eqref{eq:tcool}, and
\begin{equation}
 L_\text{cool} < \epsilon_\text{edd}L_\text{edd}
\end{equation}
where $L_\text{edd}$ is the Eddington luminosity of the central galaxy's black hole.

\subsubsection{Disc stability}
\label{sec:stability}

Discs are considered stable if they satisfy
\begin{equation}
\frac{V_\text{max}}{\sqrt{{1.68} \,G M_\text{disc}/r_\text{disc}}}<  \fstab
\end{equation}
where $\fstab$ is a model parameter close to 1. If at any timestep this
criterion is not satisfied, it is assumed that the disc is quickly converted
into an spheroid due to a disc instability and a starburst is triggered
\green{-- i.e. all the gas and stars are instantaneously moved into the spheroid
component where the star formation follows equation~\eqref{eq:burst}.}

\section{Bayesian Emulation Methodology}
\label{sec:emulator}

The use of complex simulation models, such as \glf, is now widespread across many
scientific areas.
Slow simulators with high dimensional input and/or output spaces give rise to several
major problems, the most ubiquitous being that of
matching the model to observed data, and the subsequent global parameter
search that such a match entails.

The general area of Uncertainty Analysis has been developed within the Bayesian
statistical community to solve the corresponding problems associated with slow
simulators~\citep{Craig97_Pressure,Kennedy01_Calibration}. A core part of this area
is the use of emulators: an emulator is a
stochastic function that mimics the \glf model but which is many orders of
magnitude faster to evaluate, with specified prediction uncertainty that varies
across the input space \citep{OHagan06_Tutorial,Vernon2010,galf_stat_sci}.
Any subsequent calculation one wishes to do with \glf can
instead be performed far more efficiently using an emulator \citep{Higdon09_Coyote2}.
For example, an emulator can be used within an MCMC algorithm to greatly speed up
convergence \citep{Kennedy01_Calibration,Higdon04_prediction, Henderson:2009aa}.
This is especially useful as for scenarios possessing moderate to high numbers of
input parameters,
MCMC algorithms often require vast numbers (billions, trillions or more) of model
evaluations to adequately explore the input space and
reach convergence: see for example the excellent discussion in
\cite{geyer2011introduction}. Such numbers of evaluations are clearly
impractical for models that possess substantial run time, such as \glf. Another
major issue with MCMC is that of pseudo convergence:
an MCMC algorithm may after a large number of iterations {\it appear} to have
converged and hence pass every convergence test, but continued running would
eventually reveal a sudden and substantial  change in chain location, showing that
the chain had not in fact reached equilibrium at all \citep{geyer2011introduction}.

Hence, although we fully support the Bayesian paradigm, we do not use an MCMC
algorithm here, due both to the reasons discussed above, and to the fact that a
Bayesian MCMC approach requires a full joint probabilistic specification across all
uncertain
quantities, that is often hard to make and hard to justify.
We instead outline a more efficient and robust approach known as iterative history
matching using Bayesian emulation~\citep{Vernon2010}. Here the set of all inputs
corresponding to acceptable matches to the
observed data is found, by iteratively removing unacceptable regions of the input
space in waves. History matching naturally incorporates Bayesian emulation and has
been
successfully employed in a range of scientific disciplines
including galaxy formation~\citep{Vernon2010,Vernon10_CS_rej,Bower2010,
galf_stat_sci}, epidemiology~\citep{Yiannis_HIV_1,Yiannis_HIV_2}, oil reservoir
modelling~\citep{Craig96_Pressure,Craig97_Pressure,JAC_Handbook,JAC_sma_samp},
climate modelling~\citep{Williamson:2013aa}
and environmental science~\citep{asses_mod}. History matching can be viewed as a
useful precursor to a fully Bayesian analysis
that is often in itself sufficient for model checking and model development.
Here we use it within a Bayes Linear framework, a simpler, more tractable
version of Bayesian statistics, where only expectations, variances and covariances
need to be
specified \citep{Goldstein_99,Goldstein07_BayesLinearBook}. However, if one is
committed to  a full Bayesian MCMC approach, performing an a priori history match
can dramatically improve the subsequent efficiency of the MCMC by first removing the
vast regions of input parameter space that would have extremely low
posterior probability.

\subsection{Emulator Construction}
\label{sec:emconstr}

We now outline the core emulator methodology \citep[see][for
further description]{Vernon2010,Bower2010}.
We represent the \glf model as a function $f(x)$, where
$x=~\!(\nusf,P_\text{sf}/k_\text{B},\dots,\epsilon_\text{strip},\alpha_\text{rp})$
is a
vector
composed of the 20 input parameters given in table~\ref{tab:parameters}, and $f$ is a
vector containing all \glf outputs of interest, specifically the GSMF at various
mass bins and redshifts. To construct an emulator we generally perform an initial
space filling set of wave 1 runs, using a maximin latin hypercube design over the
full 20 dimensional input space
\citep[see][for details]{Bower2010,SWMW89_DACE,Santner03_DACE,Currin91_BayesDACE}.
For each output $f_i(x)$, $i=1\dots q$, a Bayesian emulator can be structured as
follows:
\begin{equation}
\label{eq_emulator}
f_i(x) = \sum_j \beta_{ij}  g_{ij}(x_{A_i}) + u_i(x_{A_i}) + v_i(x)
\end{equation}
Here $\beta_{ij}$, $u_i(x_{A_i})$ and $v_i(x)$ are uncertain quantities to be
informed by the current set of runs.
The active variables $x_{A_i}$ are a subset of the inputs that are found to be most
influential for output $f_i(x)$. The $g_{ij}$ are known deterministic
functions of $x_{A_i}$, with a common choice being low order polynomials, and the
$\beta_{ij}$ are unknown regression coefficients. $u_i(x_{A_i})$ is a Gaussian
process with, for example, zero mean and possible covariance function:
\begin{equation}
\label{eq_corr}
{\rm Cor}(u_i(x_{A_i}),u_i(x'_{A_i})) = \sigma^2_{u_i} {\rm exp}\left\{-
\|x_{A_i}-x_{A_i}'\|^2 / \theta_i^2 \right\}
\end{equation}
where $\sigma^2_{u_i}$ and $\theta_i$ are the variance and correlation length of
$u_i(x_{A_i})$ which must be specified,
and $v_i(x)$ is an uncorrelated nugget with expectation zero and ${\rm Var} (v_i(x))
=  \sigma^2_{v_i}$, that represents the effect of the remaining inactive input
variables, and/or any stochasticity exhibited by the model \citep{Vernon2010}.

We could employ a fully Bayesian approach by specifying joint prior distributions for
all uncertain quantities in equation~(\ref{eq_emulator}), and subsequently updating
beliefs about $f_i(x)$ in light of the wave~1 runs via Bayes theorem.
Here instead we prefer to use the more tractable Bayes Linear approach, a version of
Bayesian statistics that requires only expectations, variances and covariances for
the prior specification, and which uses only efficient matrix calculations, and no
MCMC
\citep{Goldstein07_BayesLinearBook}.
Therefore if we are prepared to specify ${\rm E}(\beta_{ij})$, ${\rm
Var}(\beta_{ij})$, $\sigma^2_{u_i}$, $\sigma^2_{v_i}$ and $\theta_i$, we can obtain
the
corresponding Bayes Linear priors for $f_i(x)$ namely ${\rm E}(f_i(x)), {\rm
Var}(f_i(x))$ and
${\rm Cov}(f_i(x),f_i(x'))$ using equations~(\ref{eq_emulator}) and (\ref{eq_corr}).

The initial wave of $n$ runs is performed at input locations $x^{(1)},
x^{(2)},\dots,x^{(n)}$ which give model output values $D_i = (f_i(x^{(1)}),
f_i(x^{(2)}),\dots,f_i(x^{(n)}))$, where $i$ labels the model output. We obtain the
Bayes Linear adjusted expectation ${\rm E}_{D_i}(f_i(x))$ and variance ${\rm
Var}_{D_i}(f_i(x))$ for $f_i(x)$ at new input point $x$ using:
\begin{align}
 {\rm E}_{D_i}(f_i(x)) =&\; \label{eq_BLE}
{\rm E}(f_i(x)) \nonumber\\
&+\, {\rm Cov}( f_i(x), D_i) {\rm Var}(D_i)^{-1} (D_i - {\rm E}(D_i))
\\
{\rm Var}_{D_i}(f_i(x)) =&\;  \label{eq_BLV}
{\rm Var}(f_i(x)) \nonumber\\
            &-\, {\rm Cov}( f_i(x), D_i) {\rm Var}(D_i)^{-1} {\rm Cov}(D_i,f_i(x))
\end{align}
The emulator thus provides a prediction ${\rm E}_{D_i}(f_i(x))$ for the behaviour of
the \glf model at new input point $x$ along with a corresponding $x$ dependent
uncertainty ${\rm Var}_{D_i}(f_i(x))$. It is the later feature that strongly
contributes to emulators being more advanced than interpolators. These two quantities
${\rm E}_{D_i}(f_i(x))$ and ${\rm Var}_{D_i}(f_i(x))$ are used directly in the
implausibility measures that form the basis of the global parameter search described
below.

\subsection{Simple 1-dimensional Example}
\label{sec:1demul}

To clarify the above description we outline the construction of a simple
1-dimensional emulator of the function
\begin{equation}
f(x) \;\;=\;\; 3 \,x \sin\left(\frac{5\upi(x-0.1)}{ 0.4}\right)
\end{equation}
for which we perform a set of $n=10$ equally spaced wave 1 runs at locations
$x^{(j)}=0.1,\dots,0.5$ giving rise to run data
\begin{equation}\label{eq_Di_sim}
D = (f(x^{(1)}), f(x^{(2)}),\dots,f(x^{(n)}))
\end{equation}
where we have dropped the $i$ subscript as the output is only 1-dimensional.

For simplicity we reduce the emulator's regression terms $\beta_{ij} g_{ij}(x_A)$, in
equation~(\ref{eq_emulator}), to a constant $\beta_0$ and remove the
nugget $v_i(x)$ as there are no inactive inputs. The emulator
equation~(\ref{eq_emulator}) therefore reduces to:
\begin{equation}\label{eq_sim_em}
f(x) \;\;=\;\; \beta_0 + u(x)
\end{equation}
A possible prior specification is to treat the constant or mean term $\beta_0$ as
known, with ${\rm E}(\beta_0)=0.1$ and hence ${\rm Var}(\beta_0)=0$.
We also set $\sigma_{u}=0.6$ and $\theta = 0.06$: a choice that represents curves of
moderate
smoothness.
We can now calculate all terms on the rhs of equations~(\ref{eq_BLE}) and
(\ref{eq_BLV}) using equations~(\ref{eq_sim_em}), (\ref{eq_corr}) and
(\ref{eq_Di_sim}), for example:
\begin{eqnarray}
{\rm E}(f(x)) &=& \beta_0 \\
{\rm Var}(f(x)) &=& \sigma_{u}^2 \\
{\rm E}(D) &=& (\beta_0, \dots, \beta_0)^T
\end{eqnarray}
while ${\rm Cov}( f(x), D)$ is now a row vector of length $n$ with $j$th component
\begin{eqnarray}
{\rm Cov}( f(x), D)_j &=& {\rm Cov}( u(x), u(x^{(j)})) \\
&=& \sigma^2_{u} {\rm exp}\left\{- \|x-x^{(j)}\|^2 /  \theta^2 \right\} \nonumber
\end{eqnarray}
and ${\rm Var}(D)$ is an $n\times n$ matrix with $(j,k)$ element
\begin{eqnarray}
{\rm Var}(D)_{jk} &=& {\rm Cov}( u(x^{(j)} ), u(x^{(k)})) \\
&=& \sigma^2_{u} {\rm exp}\left\{- \|x^{(j)}-x^{(k)}\|^2 /  \theta^2 \right\}
\nonumber
\end{eqnarray}
We can now construct the emulator by calculating the adjusted expectation and
variance ${\rm E}_{D}(f(x))$ and ${\rm Var}_{D}(f(x))$ from equations~(\ref{eq_BLE})
and (\ref{eq_BLV}) respectively, for any new input point $x$.

\begin{figure}
 \centering
 \includegraphics[width=\columnwidth]{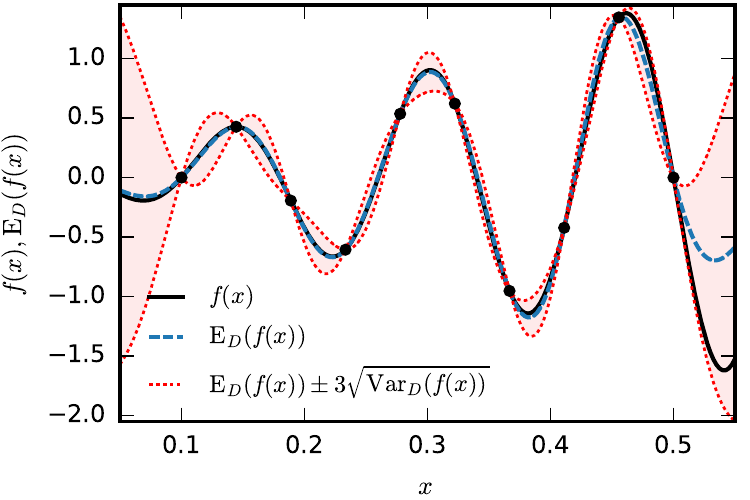}
 \caption{
The 1-dimensional emulator as constructed in \S\ref{sec:1demul}. The
\green{dashed (blue)}
line is the emulator prediction ${\rm E}_{D}(f(x))$ as a function of $x$, and
the credible
interval ${\rm E}_{D}(f(x)) \pm 3 \sqrt{{\rm Var}_{D}(f(x))}$ is given by the
\green{dotted (red)}
lines. The true function $f(x)$ is shown as the black solid line, and the 10 model
runs that make up the vector $D$ used to build the emulator are given as the red
points.
         }
 \label{fig:1demul}
\end{figure}

Fig.~\ref{fig:1demul} shows the 1-dimensional emulator where ${\rm E}_{D}(f(x))$ as
a function of $x$ is given by the \green{dashed} blue line, and the credible interval
${\rm E}_{D}(f(x)) \pm 3 \sqrt{{\rm Var}_{D}(f(x))}$ by the \green{dotted} red lines. We can see
that ${\rm E}_{D}(f(x))$ precisely interpolates the known runs at outputs $D$, with
zero uncertainty (as the red lines touch at these points): a desirable feature as
here $f(x)$ is a deterministic function. The
credible regions get wider the further we are from known runs, appropriately
reflecting our lack of knowledge in these regions.
The true function $f(x)$ is given by the solid black line which lies within the
credible region for all $x$, only getting close to the boundary for $x>0.5$.
This demonstrates the power of an emulator: using only a small number of runs we can
successfully mimic relatively complex functions to a known accuracy, a feature that
scales well in higher dimensions due to the chosen form of the emulator. The speed of
Bayesian emulators is also crucial
for global parameter searches where we may need to evaluate the emulator a huge
number of times to fully explore the input space.
Note that the
emulator calculation is extremely fast because it only requires matrix multiplication
for each new $x$. The inverse ${\rm Var}(D_i)^{-1}$ that features in
equations~(\ref{eq_BLE}) and (\ref{eq_BLV}) is independent of
$x$ (and indeed of $D_i$) and hence can be performed only once, offline and in
advance of even the run evaluations $D_i$.

\subsection{Emulating in Higher Dimensions}\label{sec:em_high_dim}

When emulating functions possessing high input dimension, the polynomial regression
terms
$\beta_{ij} g_{ij}(x_{A_i})$ in the emulator equation~(\ref{eq_emulator}) become more
important, as they efficiently capture many of the more global features often present
in the physical model~\citep{Vernon2010,Vernon10_CS_rej}. Prior specifications for
the $\beta_{ij}$ can be given,
based say on structural knowledge of the model, or on past experience running a
faster but simpler previous version of the model \citep{JAC_Handbook}. However, if no
strong prior
knowledge is available and the number of runs performed is reasonably high, a vague
prior limit can be taken in the Bayes Linear update equations~(\ref{eq_BLE}) and
(\ref{eq_BLV}), resulting in the adjusted expectation and variance of the
$\beta_{ij}$ terms tending toward their Generalised Least Squares (GLS) estimates.
For space filling runs, such as those from a maximin latin hypercube, the GLS
estimates can be accurately approximated by the
corresponding Ordinary Least Squares (OLS) estimates, which can also be used to
estimate
$\sigma_{u_i}^2$, providing further efficiency gains \citep{Vernon2010}.

In addition, the choice of active input variables $x_{A_i}$ and the choice of the
specific regression terms $\beta_{ij} g_{ij}(x_{A_i})$ that feature in the emulator,
can both be made using linear model selection techniques based on AIC or BIC
criteria. For example, these can be simply employed using the lm() and step()
functions in R \citep{Vernon2010,R-Core-Team:2015aa}. The use of active variables
$x_{A_i}$ can lead to substantial
dimensional reduction of the input space of each of the outputs, and hence convert a
high dimensional
problem into a collection of low dimensional problems, which is often far easier to
analyse \citep[see][for further discussion of this benefit]{Vernon10_CS_rej}. It is
worth noting that reasonably accurate emulators can often be constructed just using
such regression models. This can be a sensible first step
\cite[see][]{Yiannis_HIV_3}, before one attempts the construction of a full emulator
of the form given in equation~\eqref{eq_emulator}.

\subsection{Iterative History Matching via Implausibility}
\label{sec:HM}

We now describe the powerful iterative global search method known as History
Matching \citep{Craig96_Pressure,Craig97_Pressure}, which
naturally incorporates the use of Bayesian emulators, and which has been
successfully applied across a variety of scientific disciplines.
It aims to identify the set  $\mathcal{W}$ of all inputs $x$ that would give rise
to an acceptable match between the \glf outputs $f(x)$ and the corresponding vector
of observed data $w$, and proceeds iteratively, discarding regions of input space
which are deemed {\it implausible} based on information from the emulators. For more
detail on the contents of this section see
\citep{Vernon2010,Vernon10_CS_rej}.

For an output $f_i(x)$ we define the implausibility measure:
\begin{equation}\label{eq_imp1}
I_i^2(x,w_i) \;\;=\;\; \frac{({\rm E}_{D_i}(f_i(x)) - w_i)^2}
{ {\rm Var}_{D_i}(f_i(x)) + \sigma^2_{\epsilon_i} + \sigma^2_{e_i} }
\end{equation}
which takes the distance between the emulator's prediction of the $i$th output
${\rm E}_{D_i}(f_i(x))$ and the actual observed data $w_i$ and standardises
it with respect to the variances of the three major uncertainties: the emulator
uncertainty ${\rm Var}_{D_i}(f_i(x))$, the model discrepancy $\sigma^2_{\epsilon_i}$
and the observation error $\sigma^2_{e_i}$.

The least familiar of these is the model discrepancy $\sigma^2_{\epsilon_i}$ which
is an upfront acknowledgement of the deficiencies of the \glf model
in terms of assumptions used, missing physics and simplifying approximations.
In addition to ensuring the analysis is more meaningful, this term guards against
overfitting, and the subsequent technical and robustness problems this can cause for
a global parameter search. See
\cite{Kennedy01_Calibration,Brynjarsdottir:2014aa,Goldstein09_Reify} for extended
discussions
on this point\footnote{It is worthwhile noting that any analysis that does not
include a model discrepancy is only meaningful given that ``the model $f(x)$ is a
{\it precise} match to the real Universe for some input $x$", and all conclusions
derived from such an analysis should be written with this conditioning statement
attached.}.
The form of the implausibility comes from the ``best input approach" which models
the link between the \glf model evaluated at its best possible input $x^*$ and the
real Universe $y$ as $y=f(x^*) + \epsilon$, where $\epsilon$ is a random quantity
representing the model discrepancy with variance $\sigma^2_{\epsilon}$, and assumes
that the observed data $w$ is measured with uncertain error $e$ with variance
$\sigma^2_{e}$, such that $w=y+e$. See
\cite{Craig97_Pressure,Vernon2010,Vernon10_CS_rej} for further justifications and
discussions.

Most importantly, a large value of the implausibility $I_i(x,w_i)$ for any output
implies that the point $x$ is unlikely to yield an acceptable match between $f(x)$
and $w$ were we to run the \glf model there, hence $x$ is deemed {\it implausible}
and can be discarded from further analysis. We therefore impose cutoffs of the form
$I_i(x,w_i) < c$ to rule out regions of input space, where the choice of $c$ is
motivated from Pukelsheim's 3-sigma rule\footnote{Pukelsheim's 3-sigma rule is the
powerful, general, but underused
result that states for any continuous unimodal distribution, 95\% of the probability
must lie within $\mu\pm 3\sigma$, regardless of its asymmetry or skew.}
\citep{threesigma}.
We can combine the implausibility measures from several outputs in various ways e.g.
\begin{equation}\label{eq_maximp}
I_M(x,w) \;\;=\;\; \max_{i \in Q} I_i(x,w_i)
\end{equation}
where $Q$ represents the subset of outputs currently considered (often we will only
emulate a small subset of outputs in early iterations). We may use the second or
third maximum implausibility instead for robustness reasons, or use multivariate
implausibility
measures to incorporate correlations \citep{Vernon2010,Vernon10_CS_rej}.

History matching proceeds iteratively, discarding implausible regions of the input
parameter space in waves.
At the $k$th wave we define the current set of non-implausible input points as
$\mathcal{W}_k$ and the set of outputs that have so far been considered for emulation
as~$Q_k$. We proceed according to the following algorithm:
\begin{itemize}
\item[]
\vspace{-0.1cm}
\begin{enumerate}[1.]
\item Design and evaluate a space filling set of wave $k$ runs over the current
non-implausible space $\mathcal{W}_k$. \label{alg:1}
\item Check if there are informative outputs that can now be emulated accurately
(that were difficult to emulate in previous waves) and add them to $Q_k$, to define
$Q_{k+1}$.
\item Use the wave $k$ runs to construct new, more accurate emulators defined only
over the region $\mathcal{W}_k$ for each output in $Q_{k+1}$.
\item Recalculate the implausibility measures $I_i(x,w_i)$, $i \in Q_{k+1}$, over
$\mathcal{W}_k$, using the new emulators.
\item Impose cutoffs $I_i(x,w_i) < c$ to define a new, smaller
non-implausible volume $\mathcal{W}_{k+1}$ which satisfies $\mathcal{W} \subset
\mathcal{W}_{k+1} \subset \mathcal{W}_k$.
\item Unless:
    \begin{enumerate}[A)]
        \item the emulator variances ${\rm Var}_{D_i}(f_i(x))$
         are now small in comparison to the other
        sources of uncertainty: $\sigma^2_{\epsilon_i} +
\sigma^2_{e_i}$,\label{alg:6a}
        \item the entire input space has been deemed implausible or
        \item computational resources have been exhausted,
    \end{enumerate}
    return to step~\ref{alg:1}.
\item If \ref{alg:6a} is true, generate a large number of acceptable runs from the
final non-implausible volume $\mathcal{W}$, using appropriate sampling for the
scientific purpose.\label{alg:7}
\end{enumerate}
\vspace{0.3cm}
\end{itemize}
We are then free to analyse the structure of the non-implausible volume $\mathcal{W}$
and the behaviour of model evaluations from different locations within it. The
history matching approach is powerful for several reasons:
\begin{itemize}
\item As we progress through the waves and reduce the non-implausible volume, we
expect the function $f(x)$ to become smoother, and hence to be more accurately
approximated by the regression part of the emulator $\beta_{ij} g_{ij}(x_{A_i})$
(which is often composed of low order polynomials -- see
equation~\ref{eq_emulator}).
\item At each new wave we have a higher density of points in a smaller volume,
therefore the emulator's Gaussian process term $u_i(x_{A_i})$ will be more effective,
as it depends mainly on the proximity of $x$ to the nearest runs.
\item In later waves the previously strongly dominant active inputs $x_{A_i}$ from
early waves will have had their effects curtailed, and hence it will be easier to
select additional active inputs, unnoticed before.
\item There may be several outputs that are difficult to emulate in early waves
(often due to their erratic behaviour in scientifically uninteresting parts of the
input space) but
simple to emulate in later waves, once we have restricted the input space to a much
smaller and more physically realistic region.
\end{itemize}
History matching can be viewed as the appropriate analysis suitable for model
investigation, model checking and model development.
Should one wish to perform a fully Bayesian analysis using say MCMC, history
matching can be used as a highly effective precursor to such a calculation in order
to rule out vast regions of input space that would only contain extremely low
posterior probability.
However such an MCMC analysis would only be warranted assuming one is willing to
specify meaningful joint probability distributions over all uncertain quantities
involved, in
contrast to only the expectations, variances and covariances required for the Bayes
Linear history
match.

\subsection{Application of Emulation and History Matching to \glf and the GSMF}

We now apply the above Bayesian emulation and history matching methodology to \glf
and the
GSMF, and generalise it to the case of multiple available observed data sets.
We first identify the \glf model outputs $f_i(x)$ that we wish to emulate, and the
corresponding observed data $w_i^{(m)}$ to match them to as
\begin{equation}\label{eq_f_and_w}
f_i(x) = \log\phi_{i,\text{model}} \quad \text{and} \quad w_i^{(m)} =
\log\phi_{i,\text{obs}_{(m)}}
\end{equation}
where
\begin{equation}
 \phi_i = \left.\frac{\dv n}{\dv \log M_\star} \right|_{M_{\star,i}, z}\nonumber
\end{equation}
is the GSMF at the stellar mass bin $M_{\star,i}$ for redshift $z$. Here $m$ labels
the choice of observed data sets we use, represented for output $i$ by
 $w_i^{(m)}$.
{Following the discussion in \citet{Bower2010}, we adopt a model
discrepancy of 0.1 dex.  This term summarises the accuracy we expect for the model
due to the approximations inherent in the semi-analytic method}.
{In effect, this means that we will regard models that lie within 5\% of the
observed data-point as a perfectly adequate fit, even if the quoted Poisson
observational errors are substantially smaller. This means that if a model has a
marginally acceptable implausibility, $\,I\sim 3$, it may be 0.3 dex away from the
observational data-point.
}

As we have multiple sets of observed data for the GSMF which we wish to match to, we
have to make an additional decision as to how to combine these within the history
matching process. Here we generalise the implausibility measure of
equation~\eqref{eq_maximp} by minimizing over the $m$
 data sets:
\begin{equation}\label{eq_maximp2}
I_M(x,w) \;\;=\;\; \max_{i \in Q} \{ \min_m  I_i(x,w_i^{(m)}) \}
\end{equation}
with the second and third maximum implausibilities defined similarly.
This implies that our history match search will attempt to find all inputs that lead
to matches to {\it any} of the observed data sets, judged on an individual bin basis.
This is
a simple
way of incorporating several (possibly conflicting) data sets into the history match
that does not involve additional assumptions or further statistical modelling,
and which is sufficient for our current purposes.
It should lead to the identification of all inputs of interest, subsets of which (for
example those that match a specific data set, or a combined data set) can be
subsequently explored in further detail.

The emulators used in each wave were constructed following the techniques described
in \S\ref{sec:emconstr}, \S\ref{sec:1demul} and specifically the high
dimensional approaches of \S\ref{sec:em_high_dim}.

\subsection{Observational datasets used}
\label{sec:obs}

For the local Universe GSMF, we use the results of \citet{Li2009} based on SDSS and
\citet{Baldry2012} on the GAMA survey.

For larger redshifts, we combine the results of \citet{Tomczak2014} based on the
\textsc{zfourge} and \textsc{candels} surveys, and \citet{Muzzin2013}, based on the
\textsc{UltraVISTA} survey.
In these papers, the GSMF is reported for redshift
intervals/bins. For simplicity, we adopt the midpoint of each redshift bin as
the typical redshift to be compared with the model (e.g. the GSMF obtained for
$0.5<z<1.0$ will be compared with the model results for $z=0.75$).
\green{Both datasets obtain their stellar masses using the \textsc{fast} code
\citep{Kriek2009} to fit the stellar  population synthesis model of
\citet{BruzualCharlot2003} to the measured spectral energy distributions of the
galaxies, assuming a \cite{ChabrierIMF} IMF.
}
\green{
Errors in the determination of galaxy masses at $z>0$ redshifts were assumed to
follow the redshift-dependent estimate as \citet{Behroozi2013}, i.e.
$\sigma_M(z) = \sigma_0 + \sigma_z z$, with $\sigma_0=0.07$ and $\sigma_z=0.04$.
These mass errors were accounted for convolving the model GSMF with a Gaussian kernel
(see \S\ref{sec:high_z} for a discussion).
}

\begin{table}
\begin{center}
\caption{
\label{tab:cutoffs}Thresholds used for eliminating implausible regions with respect
to the local Universe GSMF after each wave and the fraction of the initial volume in
the non-implausible region.}
\begin{tabular}{ccccc}
\hline
Wave     & \mc{3}{c}{Threshold}    &   Fraction of the               \\
& 1$^{\rm st}$ max. & \mc{1}{c}{2$^{\rm nd}$} max.& \mc{1}{c}{3$^{\rm rd}$} max.&
initial volume \\
\hline
1 & -    & \mc{1}{c}{3.2}  & \mc{1}{c}{2.5} & 0.2522\\
2 & 4.5  & \mc{1}{c}{3.0}  & \mc{1}{c}{2.3} & 0.0494\\
3 & 3.75 & \mc{1}{c}{2.5}  & \mc{1}{c}{2.0} & 0.0170\\
4 & 3.5  & \mc{1}{c}{2.5}  & \mc{1}{c}{2.0} & 0.0116\\
5 & 3.0  & \mc{1}{c}{2.25} & \mc{1}{c}{2.0} & 0.0036\\
6 & 2.4  & \mc{1}{c}{2.15} & \mc{1}{c}{1.8} & 0.0010\\
\hline
\end{tabular}
\end{center}
\end{table}

\section{Results}

\label{sec:results}

The parameter space exploration was conducted through successive waves of runs.
After each wave, emulators were generated from its results and used to design
the parameter choices for the next wave, discarding a vast, specifically
implausible, region of the parameter space. Each wave was designed using a
Latin Hypercube Sampling of $5000$ points of the non-implausible region of the
parameter space (full details are given in appendix \ref{ap:emu_details}).

\subsection{Matching the local galaxy stellar mass function}
\begin{figure}
 \centering
 \includegraphics[width=\columnwidth]{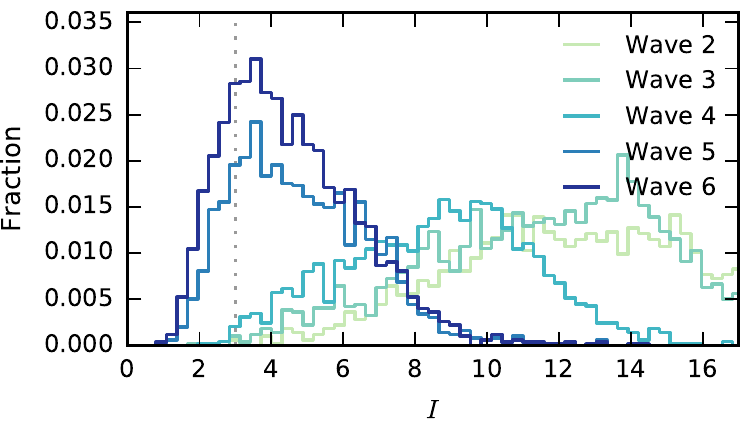}
 \includegraphics[width=\columnwidth]{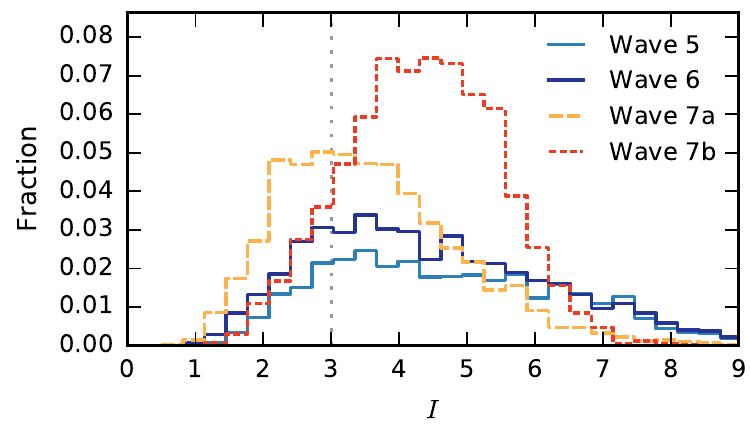}
 \caption{
          Histograms of the model implausibilities (with respect to the local
          Universe only) obtained at each wave.
          In the top panel, exploratory waves 2-6 are shown.
          Each new wave reduces the tail of very implausible models.
          However, the increase in the number of models with $I<2$ occurs only slowly
          after each wave.
          In the bottom panel, waves 7a and 7b are also shown (please, note the
          different range in $I$). Wave 7a was designed specifically to obtain many
          plausible runs, instead of uniformly covering the non-implausible
parameter space, and Wave 7b (discussed in \S\ref{sec:high_z}) takes into account
          the constraints by high redshift data.
          }
 \label{fig:wall}
\end{figure}

\label{sec:z0}
\begin{figure}
 \centering
 \includegraphics[width=\columnwidth]{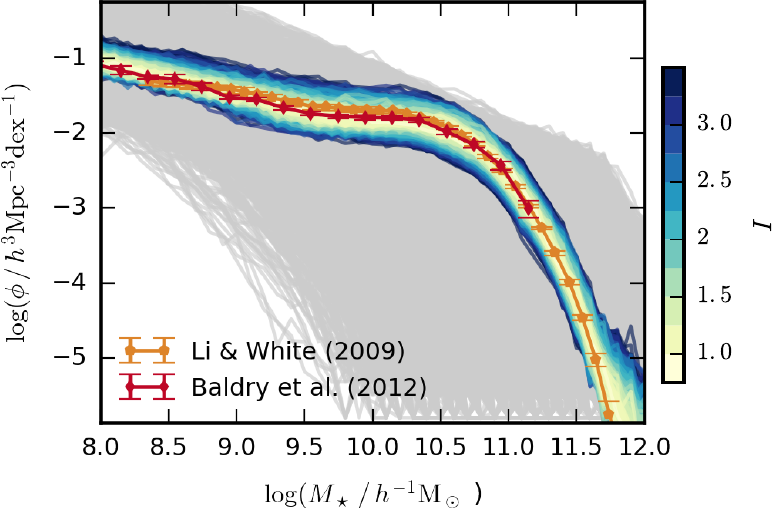}
 \caption{
          Local galaxy stellar mass function.
          The coloured curves show all the runs with implausibility $I<3.5$, with
          different shades showing different implausibilities (see colour-bar in the
          plot).
          The surrounding light grey curves correspond to the initial set of runs
          (wave 1).
          Data-points show the GSMF data obtained by \citet{Li2009} and
          \citet{Baldry2012}.
          {Note that a 0.1 dex model discrepancy was assumed (see text for
          details).}
         }
 \label{fig:smf_z0}
\end{figure}

There were initially 6 waves of runs, where the implausibilities were computed with
respect to the local Universe GSMF data only.
Table \ref{tab:cutoffs} shows the implausibility cut-off thresholds applied,
which decreased after each wave as we build more trustworthy emulators.
Table \ref{tab:cutoffs} also shows the fraction of the
initial volume which corresponds to the region classified as non-implausible after
each wave.
To compute the volumes, the parameters were rescaled following equations
\eqref{eq:scale_lin} and \eqref{eq:scale_log} -- i.e. lengths associated with the
range of each parameter were considered equivalent.
Despite making very conservative choices
for the thresholds, there is a strong decrease in the volume after each wave and
after wave 6 only $10^{-3}$ of the original volume was classified as
non-implausible.
In Fig. \ref{fig:wall} the evolution of the distribution of the
implausibilities can be followed: after each wave the number of highly
discrepant models is strongly reduced, but the number of acceptable models,
with $I<3$, increases only modestly.

\begin{figure*}
\centering
\includegraphics[width=\textwidth]{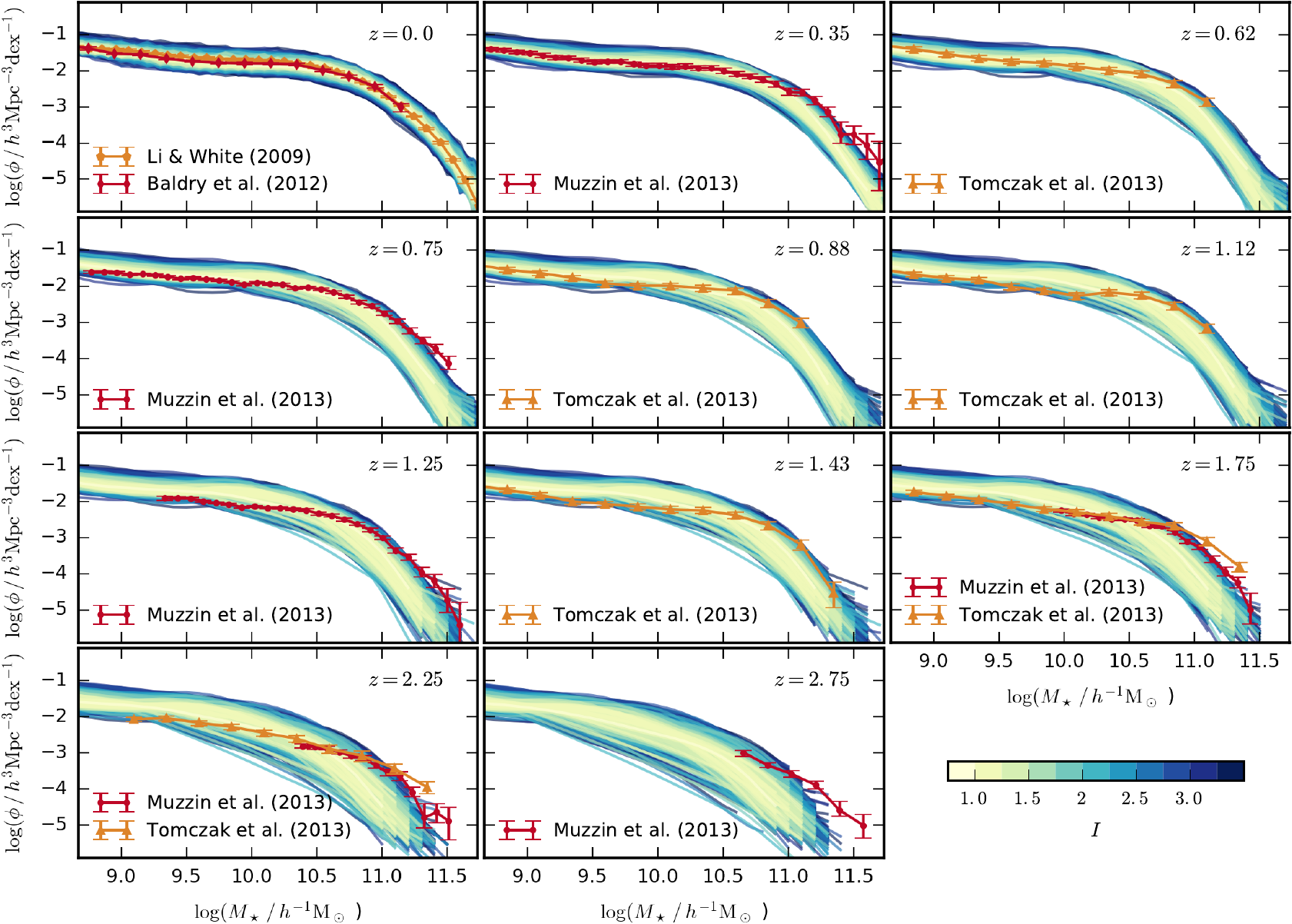}
\caption{
         The GSMF of selected models colour-coded by their
         implausibility, $I$ (see colour bar),
         calculated \emph{only} with respect to the GSMF
         data at the local Universe, from \citet{Baldry2012} and \citet{Li2009}.
         Only runs with $I<3.5$ with are shown.
         The high redshift observational data
         shown were obtained from \citet{Muzzin2013} and \citet{Tomczak2014}.
         \label{fig:smf_grid_z0}
         The models selected solely by their good match to $z=0$ data produce
         poor agreement to higher redshift data.
        }
\end{figure*}

\begin{figure*}
\centering
\includegraphics[width=\textwidth]{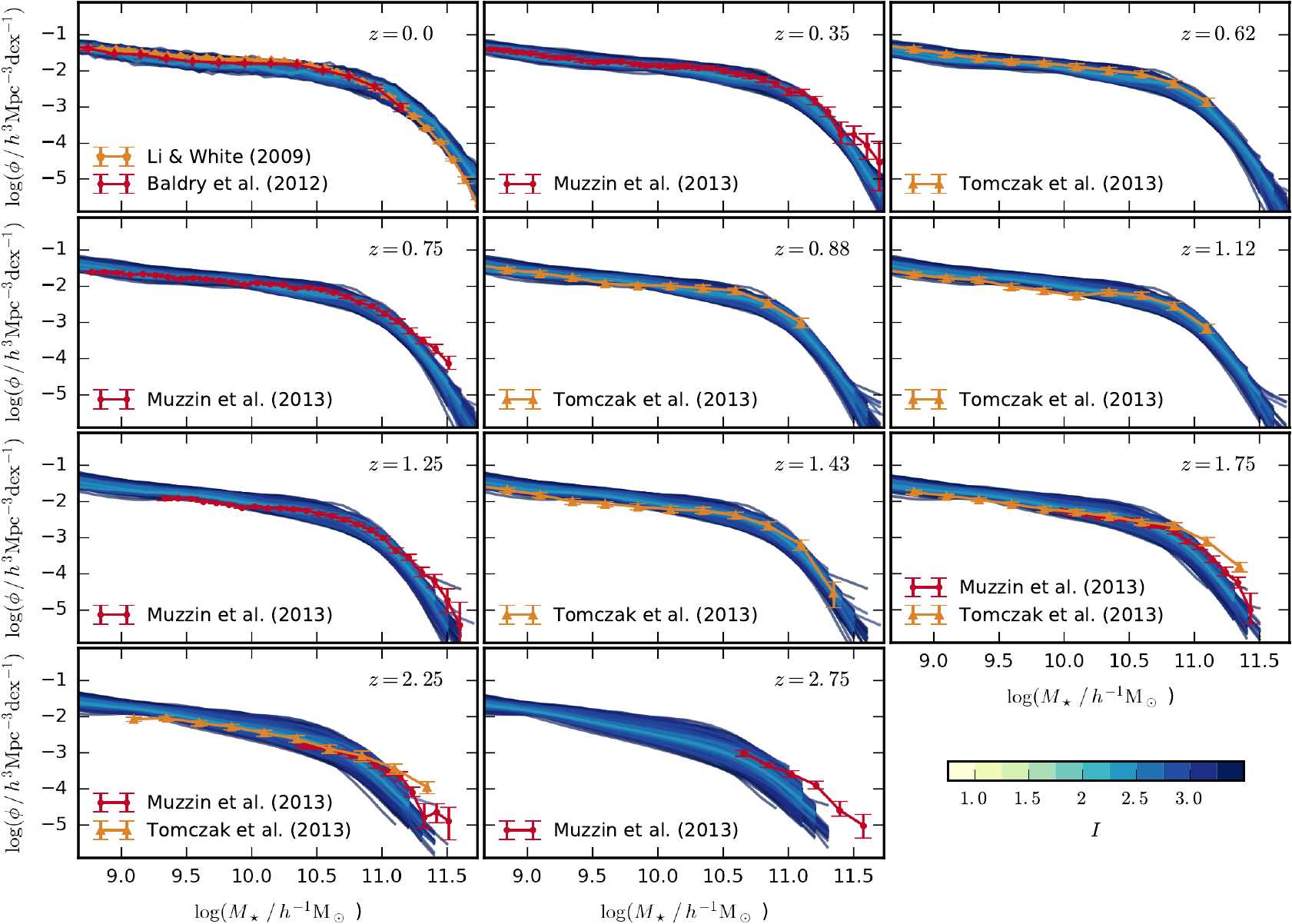}
\caption{\label{fig:smf_grid_high}
         The GSMF of selected models colour-coded by their
         implausibility, $I$, calculated \emph{simultaneously} with respect to
         the GSMF at
         $z=0,\,0.35,\,0.62,\,0.75,\,0.88,\,1.12,\,1.25$ and $1.43$.
         Only runs with $I<3.5$ are shown.
         The high redshift observational data
         shown were obtained from \citet{Muzzin2013} and \citet{Tomczak2014}
         \green{and local Universe data were obtained from \citet{Baldry2012}
         and \citet{Li2009}.}
         \glf models generally lead to a GSMF with a high mass end which is too
         shallow
         at $z=0$ and too steep at higher redshifts.
        }
\end{figure*}

After the 6$^{\rm th}$ wave the emulator variance at each point was already
smaller or equal to the other uncertainties,
indicating that no further refinement was possible (condition \ref{alg:6a}
in the algorithm described in \S\ref{sec:HM}). A final, wave 7a,
was then designed, this time using the emulator information to aim for
the best possible matches to the GSMF (i.e. step \ref{alg:7} in the algorithm;
in contrast to the previous waves where the non-implausible space was uniformly
sampled to ensure an optimum input for the next wave).

In Fig. \ref{fig:smf_z0}, the final results of the history matching are shown,
together with our observational constraints.
{Error bars in this and other figures show only the quoted
observational errors, and do not include the model discrepancy term, so that the
over all quality of the fits can be judged from figures.
The purpose of the model discrepancy is to avoid rejecting models when the
observational errors become very small.}
All the models in our library with implausibility $I<3.5$ are
shown {in Fig.~\ref{fig:smf_z0}}, with the curves
colour-coded by the implausibility.
{The width of the lightest shade, corresponding to $I\leq 1.0$,
allows one to visualize the effect of adopted model discrepancy.
}
Also shown are the wave 1 runs given as the grey curves, many of which
were far from the observed data.
The impact of the history match in terms of the removal of substantial amounts of
implausible regions of the parameter space can be seen by comparing the coloured
region with the grey curves.

While there are models with $I\sim2.5-3$ which produce an excess in the number of
small
mass galaxies, the opposite (i.e. $\phi$ smaller than the observations at the low
mass end) is very rare. A similar behaviour is also present in the high mass end.
Thus, acceptable ($I\lesssim 3$) models may display over-abundances of very small
[$\log(M_\star/\msun)\lesssim 8.5$] or very large
[$\log(M_\star/\msun)\gtrsim 11.5$] masses, but there are no acceptable models
with significant under-abundances in these ranges.

Once the locus of models with good fits to the local Universe GSMF was found, we
examined how well these models performed with respect to high redshift data. In
Fig. \ref{fig:smf_grid_z0} the GSMF output by the models shown in Fig.
\ref{fig:smf_z0} is now compared with higher redshift data.
One finds that the models selected only by their ability to
reproduce the local Universe GSMF data under-represent the abundance of high mass
galaxies at higher redshifts while simultaneously generating an excessive number of
galaxies of lower masses.

In the following  section we will examine the parameter space and show that the vast
majority of acceptable models have $\betaburst > \betadisk$ and so lie in a region
of parameter space not available to the original model.

\subsection{Constraining models with higher redshift data}
\label{sec:high_z}

\begin{figure*}
\centering
\includegraphics{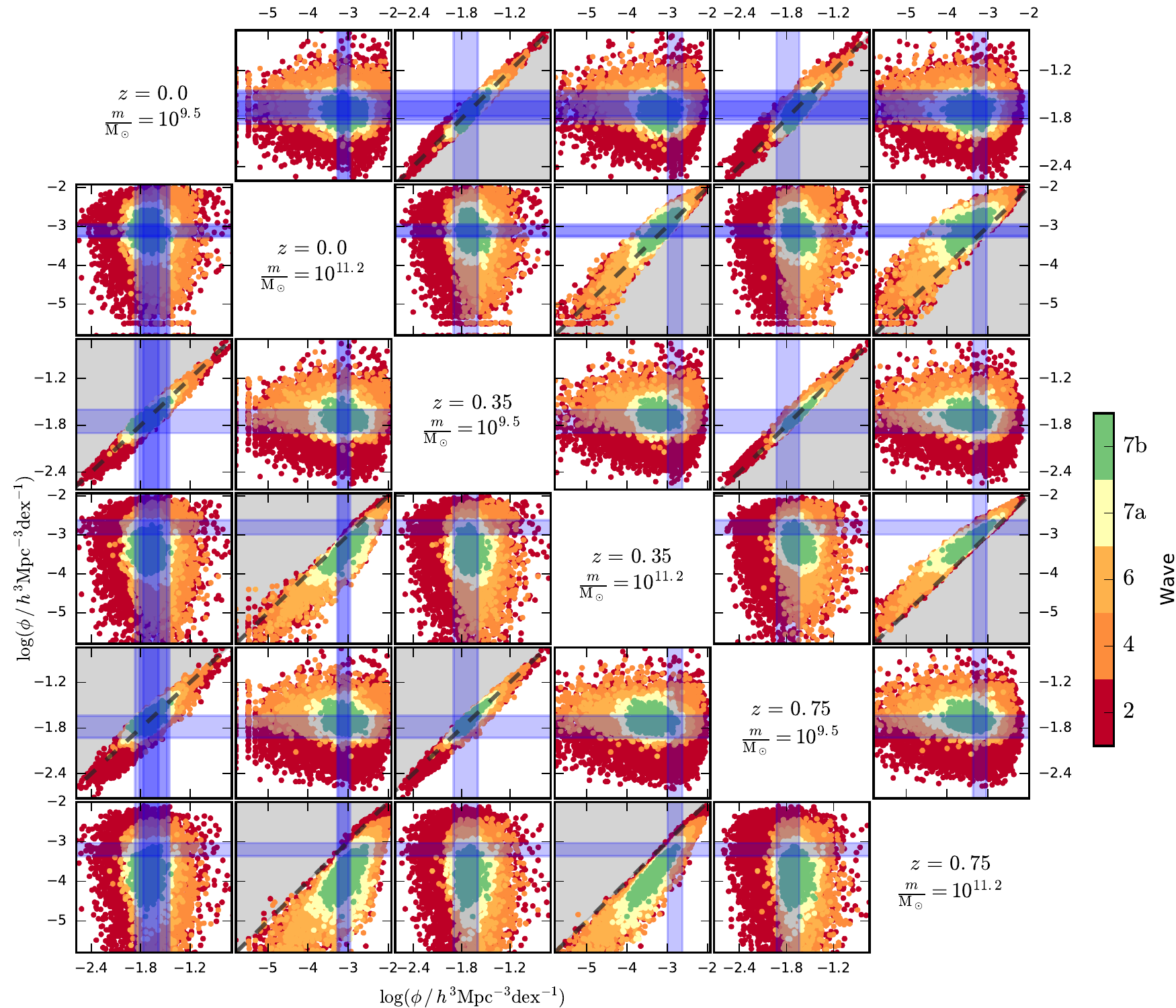}
\caption{
Each panel compares the output of the GSMF for different mass bins and/or redshifts.
Waves 2, 4, 6, 7a, 7b are shown colour-coded as indicated.
The observational constraints of previous figures are shown as blue shades.
The light grey shade was added to guide the eye, indicating where the GSMF values in
the vertical axis are smaller than the horizontal axis, and the dashed
diagonal line indicates the case where the GSMF for the two is the same.
Panels below the diagonal use the same estimates for the errors in the mass
determination as \citet{Behroozi2013}: $\sigma_M = 0.07 + 0.04\,z$ (see text for
details). For comparison, in the panels above the diagonal we double the mass error.
After successive waves there is improvement in the
agreement with the low redshift data and with the low-mass-end of the high
redshift data, however, for the high-mass end, there is still tension at higher
redshifts. The increase in the mass error does not avoid the tension.
This tension originates from the data being consistent with small or no evolution
for bins of large mass.
}\label{fig:output_pair}
\end{figure*}

To investigate if, and to what extent, the present model could reproduce the
evolution of the GSMF, a new wave of runs was generated from the wave 6 emulator
(wave 7b), this time computing the implausibilities simultaneously with respect to
the GSMF data at higher redshifts, up to $z=1.75$.

After just a single additional wave, the emulation technique indicated that no extra
refinement was likely: the emulator variances became smaller than the other
uncertainties, corresponding to step 6A in the algorithm of §3, suggesting that it
would be highly unlikely to find a locus of more plausible runs within any sub-volume
of the parameter space. A new (and final) wave was then designed, to produce runs
which provided a good match to the GSMF at those redshifts (corresponding to step 7).
This set of runs was deliberately focused towards the regions of lowest emulator
implausibility, where we would now expect the best matches to occur. This is a good
technique for exploring the correlations between parameter sets; however, it is
important to note that the resulting design of runs would not be a suitable basis for
the construction of a statistical emulator.

In Fig. \ref{fig:smf_grid_high} we show the evolution of the GSMF for all the runs
(of all waves) with implausibility $I<3.5$ with respect to redshifts up to
$z=1.12$.
The adoption of higher redshift constraints leads to tension with the local Universe
data: the least implausible models produce a GSMF with a too shallow high mass end
at $z=0$ and too steep at any other redshift. In the low mass end, there is an
excess of $\lesssim 10^{10} \msun$ galaxies at higher redshift and a small deficit
of them in the local Universe.
This is consistent with behaviour seen in the runs constrained at $z=0$ only.
{\green{It should be noted that, despite the tension,
the level agreement achieved is still better than what is found in most published models, and is not dissimilar from what is found by \citet{Henriques2013,Henriques2015}.}}

This tension becomes clearer when the results for specific mass bins of the GSMF
are compared. This is shown in panels below the diagonal in
Fig.~\ref{fig:output_pair}, for two mass bins, $\log(m/\msun h)=9.5$ and $11.2$,
redshifts $z=0.0$, $0.35$ and $0.75$. The observational constraints are shown as
blue bands. By showing the constraints in pairs, we gain insight into the conflicting
pressures imposed on the model.
Initially, successive waves of runs (shown by colours from red to green\green{,
as indicated in the figure}) are increasingly
focused towards the point at which the two bands intersect.
However, it becomes increasingly evident that some constraint pairs cannot be matched
by the model and the successive waves lead to no improvement. For example, the panel
showing $\log(m/\msun h)=11.2$ at $z=0$ and $z=0.35$ has a strong diagonal line above
which
the model is never able to cross. The same behaviour is found when comparing the high
mass end of the GSMF at $z=0$ with other redshifts. For the constraints at
$\log(m/\msun h)=9.5$,
the outputs of all models are tightly correlated when comparing between redshift.
Comparison
between different mass bins appears to be less constraining.

One can best interpret Fig.~\ref{fig:output_pair} by comparing the models to a
non-evolving GSMF.  This is shown by the dashed diagonal line in panels that
compare the same mass bins. The
grey-shaded side of the dashed diagonal line show the case where the number of
galaxies decreases with time. It can be seen that the observational data used leads
to no
evolution (or even decrease) in the number of $10^{11.2}h^{-1}\msun$ galaxies if
$z=0$ is compared to other datasets. This makes it clear why it is not possible to
find models in the exact target region. Since \glf is inherently hierarchical, it is
difficult
to conceive of a mechanism which could lead to a significant decrease in the
abundance of massive galaxies with time.  This would only be possible if
$10^{11.2}h^{-1}\msun$ galaxies
were to grow in mass (and so leave the mass bin) faster than lower mass (and more
abundant) galaxies were able to grow and move into the bin. Clearly, the situation
never arises in the \glf model and the only way of obtaining points in the grey
region for the high
mass bin panes is due to the distortion caused by errors in the galaxy mass
determination, as we will discuss below.

Systematic errors in the determination of galaxy masses (`mass errors' for short)
arising from the modelling of the star formation history, choice of dust model and
the choice of IMF can significantly affect the shape of the GSMF which is inferred
from the observations \citep{Mitchell2013}.
\green{As mentioned in \S\ref{sec:obs}, mass errors were accounted for by convolving
the model GSMF with a Gaussian kernel.
}
The main effect of such convolution is making the GSMF appear less steep at higher
redshifts.
This raises the question of whether underestimated mass errors could explain the
difficulty in simultaneously matching the high mass end of the GSMF at different
redshifts.

\bigskip

In the panels above the diagonal of Fig.~\ref{fig:output_pair}, we show
the consequences of \green{doubling the mass error -- i.e.}
considering $\sigma_0=0.14$ and $\sigma_z=0.08$.
This has the effect of loosening the implausibility contours: the
blue regions are the same as those below
the diagonal. The effect of these much increased mass errors is to
allow models near to the ``no evolution'' region, alleviating the tension by allowing
the corrected \glf
results to get closer to the target region. However, even considering these mass
errors, some tension still persists.

\subsection{Plausible models subspace}
\label{sec:subspace}
\begin{figure*}
\includegraphics[width=\textwidth]{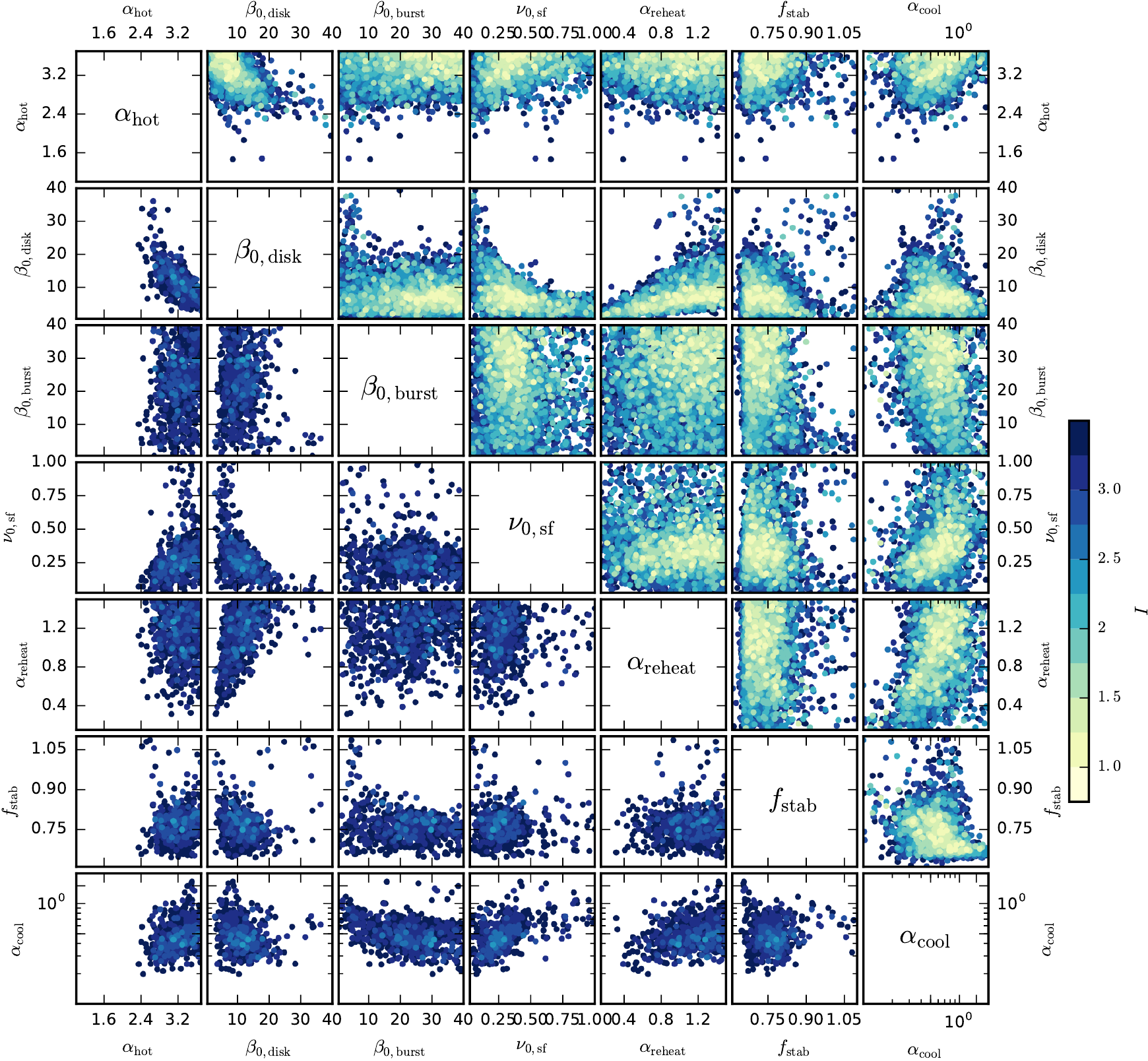}
\caption{The panels show two dimensional projections of the plausible
         parameter space. Each circle
         represents a \glf run and is colour coded by its implausibility (as
indicated by the colour bar); lower implausibility runs are plotted on
         top facilitating the visualisation of their clustering in the projected
         space; only runs with $I<3.5$ are shown.
         In panels above the diagonal, the implausibility is computed with
        respect to the observed local GSMF only.
         In panels below the diagonal, the implausibility is computed with
         respect to the GSMF at redshifts
         $z=0.0,\,0.35,\,0.62,\,0.75,\,0.88,\,1.12,\,1.25$ and $1.43$.
         Note that the axes are labelled consistently above and below the diagonal. A
panel below the diagonal should be rotated and inverted in order to compare it to the
equivalent panel above the diagonal.
         This figure summarizes the main constraints imposed by the GSMF and its
         evolution on the \glf parameters.
        }
\label{fig:implaus_space}
\end{figure*}

We examine now what are the main properties of the subspace of plausible models,
which we define as models having implausibility, $I<3.5$, a
conservative threshold.

We begin by considering the models that provide a plausible match to the
GSMF at $z=0$. The distribution of these models are shown above the diagonal
in Fig.~\ref{fig:implaus_space}.  In each panel we show the plausible models
projected
into the two dimensional space of a pair of variables.  The models are coloured by
implausibility and the lowest implausibility runs are plotted last to ensure they are
visible. This method of plotting also gives
a good impression of the ``optical depth'' of the parameter region in the hidden
parameters of each panel.
We only show the most interesting variables in this plot, the panels for other
variable pairs are less informative scatter plots.

The most constrained parameters are: the disk wind parameters, $\alphahot$ and
$\betadisk$, the normalisation of the star formation law, $\nusf$, the AGN feedback
parameters, $\alphacool$, and the disk stability threshold, $\fstab$.  Several
parameter degeneracies can be picked out in the figure.
For example, values of $\alphahot$ are strongly
correlated with $\betadisk$, with larger $\betadisk$ being compensated by a
smaller $\alphahot$: i.e., the higher mass loading normalisation is compensated by a
weaker mass dependence so that the level of feedback is similar in low-mass galaxies.

Other parameters are more weakly constrained, and it is possible to find plausible
models over most of the range of the parameter considered. The parameter
$\alphareheat$ is a good example. In this case, smaller values of $\alphareheat$ can
be compensated by reductions in $\betadisk$.
This makes physical sense. The time-scale on which gas
is re-incorporated into the halo after ejection depends on $\alphareheat^{-1}$
(equation~\ref{eq:reheat}), so that increases in the time-scale can be offset by an
overall lower mass loading of the disk wind \citep{Mitchell2016}.

One surprising feature is that the normalization mass loading associated with star
burst galaxies, $\betaburst$, (see \S\ref{sec:SN_feedback}) is weakly constrained.
Although the best models (and also the greatest number of models) have $\betaburst >
20$, entirely plausible models can be found with much
smaller values. This is presumably because the impact of the large values of
$\betaburst$ can be offset
by adjusting the values of other parameters. The pairs plot does not, however, reveal
an obvious
interaction with another individual parameter.  In \S\ref{sec:pca}, we will use a
principle
component method to try to isolate simpler interactions between parameter
combinations, and we explore the physical interpretation there.

The panels below the diagonal line show the models that generate plausible fits
to the GSMF
over the redshift range $z=0$ to $1.43$. A panel below the diagonal must be rotated
and inverted in order to compare it to the equivalent panel above the diagonal. As we
have already discussed, this is a stringent requirement, and even the best models
have $I>2$. The volume of the parameter space within which plausible models can be
found is significantly reduced compared to the situation if only the $z=0$
implausibility is considered.
The plausible range of the parameters $\alphareheat$, $\alphacool$ and $\nusf$ is
particularly affected.
For example, the addition of the high redshift GSMF excludes very long gas cycling
time-scales (and thus
small values of $\alphareheat$).

Plotting the data in this way does not, however, expose any new correlations between
parameters, or make
it easy to appreciate the physical differences in the model that result in the very
different behaviour
at high redshift that can be seen by comparing Figs. \ref{fig:smf_grid_z0} and
\ref{fig:smf_grid_high}.
In order to make it easier to identify these differences, we will analyse the
distribution of the plausible models in the Principle Component Analysis (PCA) space.
 This allows us to better identify the critical parameter combinations that are
picked out by the data. We have already noted that several parameters show
significant (anti-)correlation, and the PCA analysis will identify the
most important relations.

One of the motivations for undertaking a full parameter space exploration is the
possibility of the existence of multiple disconnected implausibility minima, which
would be unlikely to be found in the `traditional' trial-and-error approach to
choosing the parameters. Nevertheless, we find that the locus of acceptable \glf runs
is connected and there are no signs of multiple minima or other complex
shapes. Because of this, the distribution of plausible models is particularly
amenable to the PCA method.

\subsection{Principal component analysis}
\label{sec:pca}

In order to obtain greater insight into the constraints imposed by the GSMF, and in
particular the constraints imposed by the higher redshift data, we performed a
principal component analysis (PCA)
on the volume of the input parameter space containing
runs with $I<3.4$ in all the datasets at
$z=0,\,0.35,\,0.62,\,0.75,\,0.88,\,1.12,\,1.25$ and
 $1.43$, giving a set of 508 runs in total.
The PCA generates a new set of 20 orthogonal variables defined as the
eigenvectors of the
covariance matrix formed from the input parameter locations of the 508 runs, ordered
by size of eigenvalue.
Therefore the first new variable (Var~1) gives the direction which has the largest
variance in the input space,
while the last (Var~20) gives the direction with the smallest variance. Usually, PCA
is applied to find the directions with
the largest variance, but here we are precisely interested in the opposite: we wish
to learn about those directions
in input parameter space that have been most constrained by the observed data. This
analysis allows the examination of the
location of acceptable runs in the rotated (and translated) PCA space, to identify
possible hidden features, and the
transformation of the (approximately) orthogonal constraints observed in the PCA
space back on to the original parameters
to aid physical interpretation.
For example, acceptable model runs all have similar values for Var~20, Var~19 etc,
and this can be inverted to express the dependencies of the variables on one
another. It is important to note that the precise components of the PCA variables
depend on their original range (and whether the variables are normalised on to a log
or linear scale).
This can be viewed in a Bayesian sense, in that we are quantifying the increase in
knowledge about the values of the variables relative to our prior knowledge. It is
also important to bear in mind that variables with similar variance are degenerate,
and that alternative combinations of them will describe the distribution of the data
similarly well, but may have a simpler physical interpretation.

\begin{figure}
\includegraphics{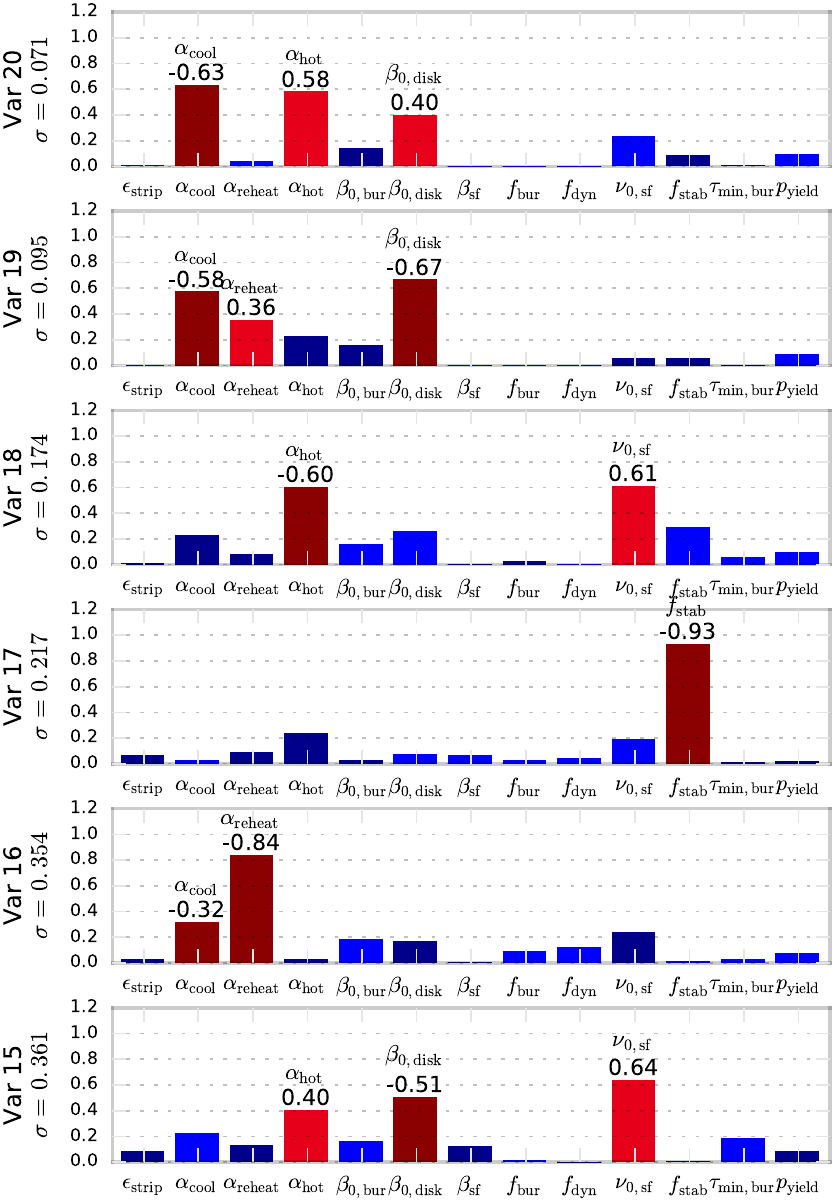}
\caption{Summary of results of the principal component analysis.
The 6 most constrained components of the region with
$I<3.4$ with respect to $z=0,\,0.35,\,0.62,\,0.75,\,0.88,\,1.12,\,1.25$ and
$1.43$ (which contains 508 models).
The bars show the absolute values of the PCA loads associated with each
scaled parameter (only parameters with non-negligible loads are shown).
Parameters with larger loads
($>0.3$)
are drawn in red
\green{and have their names and loads written on the top of the bars.}
Bright (dark)
colours show variables with positive (negative) loads.
}
\label{fig:pca}
\end{figure}

The resulting PCA variables (and the centroid of the distribution) are listed in
Appendix \ref{ap:fullPCA}.
The standard deviation in the directions defined by Var~20 and
Var~19 is extremely small (less than 0.1 relative to the prior distribution of
$\pm1$). Var~18 and Var~17 are also significantly constrained (std. dev. less than
0.22). The constraints on the other variables are much less significant, Var~14, 15
and 16 all have std. dev. $\sim 0.4$. This gives us a quantitative measure of the
information content of the GSMF relative to the freedoms of the model.

The components of the 6 most constrained variables are shown in Fig.~\ref{fig:pca}.
We begin by considering the strongly constrained components Var~19 and Var~20. The
variance of these two components is similar and so we should consider them together.
As shown by the colouring of the histogram, Var~19 is dominated by $\betadisk$ and
$\alphacool$, with a smaller contribution from $\alphareheat$. Qualitatively, this
simply confirms that the break of the GSMF is controlled by competition between AGN
and stellar feedback;
stronger winds from disks in $200\,\text{km}\,\text{s}^{-1}$ galaxies (i.e., larger
        $\betadisk$, equation~\ref{eq:beta}), or a longer re-incorporation
time-scale (i.e.,
smaller $\alphareheat$, equation~\ref{eq:reheat}), need to be compensated by an
increase the halo mass at which AGN become effective (i.e., smaller $\alphacool$,
since $t_\text{cool}/t_\text{ff}(r_\text{cool})$ increases with halo mass,
equation~\ref{eq:tcool}). As well as providing qualitative insight, this can be
translated
into quantitative constraints on the input parameters. To do this we neglect the
dependence on parameters with small loads ($<0.3$, shown in blue in
Fig.~\ref{fig:pca}) and assume that they have values close to the centroid
of the PCA expansion. Using superscripts to denote that this relation applies to
the rescaled variables (given by equations \ref{eq:scale_lin} and
\ref{eq:scale_log}),
the constraint can then be simplified to:
\begin{align}
\label{eq:var19}
|\text{Var}\,19|\, =&\,\,
|\,-0.669(\betadisk^{(s)}+0.464)
      -0.576(\alphacool^{(l)}-0.065) \nonumber\\
     & +0.356(\alphareheat^{(s)}-0.462)
     \,|\,\, \lesssim 0.095.
\end{align}

Var~20 is mainly composed of $\alphacool$ (the AGN feedback parameter), $\alphahot$
and $\betadisk$ (the quiescent feedback parameters). Eliminating variables with small
weight, we arrive at the following inequality:
\begin{align}
|\text{Var}\,20|\, =& \,\,|\,
     +0.401(\betadisk^{(s)}+0.464)
     +0.583(\alphahot^{(s)}-0.673)\nonumber\\
    &-0.634(\alphacool^{(l)}-0.065)
     \,|\,\, \lesssim \,0.071.
\end{align}
Physicaly, this relation tells us that if we pick the disk feedback parameters
$\alphahot$ and $\betadisk$, the AGN feedback must follow from the equality.
Increases in  $\alphahot$ and/or $\betadisk$ (making supernovae driven feedback) need
to be compensated by increases $\alphacool$ (making AGN feedback effective only in
higher mass haloes). Since Var~19
already determines $\alphacool$, it is more useful to write the constraint as
(neglecting small weights):
\begin{align}
|\text{Var}\,20|\, \approx& \,\,|\,
     +1.137(\betadisk^{(s)}+0.464) -0.391(\alphareheat^{(s)}-0.462) \nonumber\\
     &+0.583(\alphahot^{(s)}-0.673)
     \,|\,\, \lesssim \,0.175,
\label{eq:approx}
\end{align}
which expresses the requirement that a given choice of $\betadisk$ (and
$\alphareheat$) parameters need to be balanced by a suitable choice of circular
velocity dependence of supernova feedback, $\alphahot$.

The next two components, Var~18 and Var~17, have significantly larger variances
($\sigma=0.174$ and $0.217$, respectively). Var~17 is almost completely determined
by $\fstab$, so that successful models require a narrow range of the stability
parameter, almost independent of the other variables.
\begin{equation}
\label{eq:var17}
|\text{Var}\,17|\, =\,\,
|\,-0.931(f_{\rm stab}^{(s)}+0.362)
    \,\,\,|\,\,  \lesssim 0.217\,.
\end{equation}
Var~18 relates the star formation
efficiency $\nusf$ to $\alphahot$, the halo mass
dependence of feedback (which in turn relates to the choice of feedback parameters
$\betadisk$ and $\alphareheat$, see equation~\ref{eq:approx}):
\begin{align}
\label{eq:var18}
|\text{Var}\,18|\, &= \,\,\nonumber
|\,\,0.613(\nusf^{(s)}+0.456)\\
    &-0.604(\alphahot^{(s)}-0.673)
    \,\,\,|\,\, \lesssim 0.174.
\end{align}
Increasing the strength of feedback in small galaxies (greater $\alphahot$) requires
that star formation is made more efficient to compensate
\green{(i.e. by increasing star formation at higher mass galaxies, maintaining thus
the total amount of stars at low $z$)}.

The remaining variables are relatively weakly constrained, but have similar variance.
They provide addition constraints on the disk and AGN feedback parameters
($\alphareheat$, $\alphacool$, $\alphahot$ and $\betadisk$) and the star formation
law ($\nusf$). Although they are weakly constrained, these relations play an
important role in determining whether models successfully match the higher redshift
GSMF data as well as the $z=0$ GSMF, as we will show below.

\subsection{Effect of GSMF constraints in PCA space}
\label{sec:pca_comparison}

\begin{figure*}
\includegraphics[width=\textwidth]{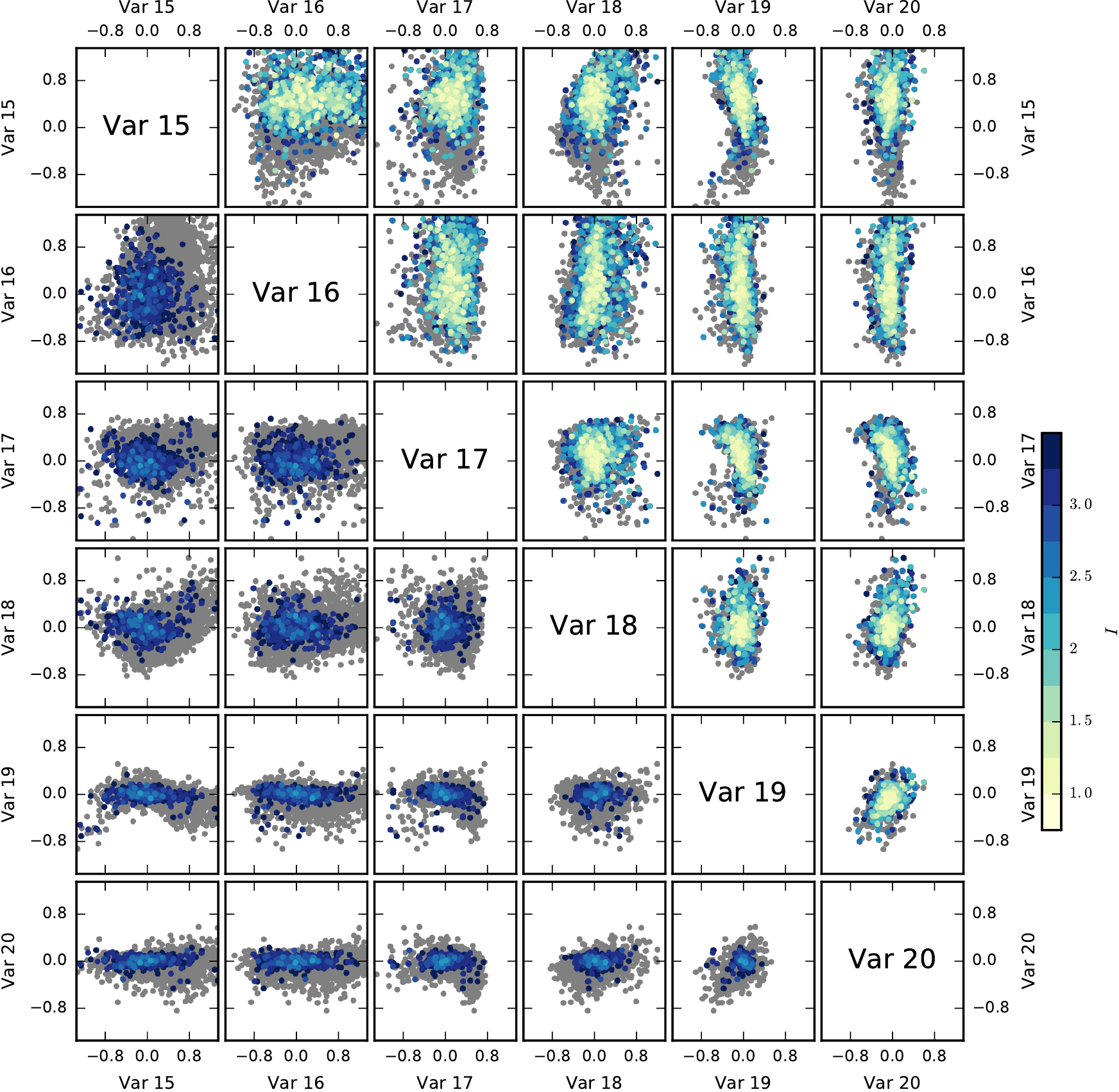}
\caption{A comparison of runs that provide plausible fits to the $z<1.43$ GSMF
datasets (below the diagonal), and those that provide a plausible description of the
$z=0$ GSMF, but a very implausible match to the high-z data (above the diagonal). We
show the comparison  PCA-space, with individual panels showing two dimensional
projections. The PCA variables are defined using the set of plausible $z<1.43$ GSMF
datasets.  We show the 6 most constrained variables (see text for discussion). Each
circle represents a \glf run and is colour coded by its implausibility (as indicated
by the colour bar); lower implausibility runs are plotted on top to facilitate the
visualisation of their clustering in the projected spaced. To facilitate comparison
of the runs above and below the diagonal, we show the full set of runs with plausibly
fits
to the $z=0$ GSMF as the underlying grey points.
        }
\label{fig:pca_pairs}
\end{figure*}

In order to better understand why some runs generate a plausible match to the
$z<1.43$ GSMF (as well as that at $z=0$) while others do not, we select the 5
components with least variance and rotate the distribution of the full set of runs
with plausible $z=0$ into this space. Note that the variables are defined using the
plausible $z<1.43$ GSMF runs, but we can use the same rotation to examine the
distribution of any set of runs. We show projections into pairs of these variables in
Fig.~\ref{fig:pca_pairs}.  Below the diagonal, we show the runs selected on the basis
of the full redshift range of GSMF data (as in Fig.~\ref{fig:implaus_space}). The
colouring, and plotting order, of points is the same as in the previous figures.
Above the diagonal, we show the set of runs that provide a good match to the $z=0$
GSMF, but a very implausible match to the full  $z<1.4$ implausibility  ($I>6$). We
add the underlying grey points to show the distribution of the runs giving plausible
fits to the $z=0$ GSMF (regardless of their $z<1.4$ implausibility) in order to make
it simpler to compare with panels above and below the diagonal.

The location of the runs in the strongly constrained variables Var~19 and Var~20
hardly changes.
These strong selection rules seem to primarily select runs with a good match to the
$z=0$ GSMF, and are not particularly important in determining whether a run also
matches the higher redshift data or not. Var~15, 16 and 17, however, show systematic
shifts above and below the diagonal, showing that it is these secondary
relationships between the feedback variables and the disk stability parameters that
are critical in matching the evolution of the mass function. In particular,
we recall that Var~17 is almost exclusively dependent on the disk stability
criterion: runs which match the $z=0$ GSMF but not the higher redshift data tend to
have higher values of Var~17, and thus lower values of $\fstab$ which tends to make
disks more unstable at low redshift.
\green{Therefore, when larger redshift data is considered, models where
instabilities are mostly present at higher redshifts are preferred.
Var~15 and 16 also show shifts, however, showing that the
re-incorporation time-scale (ie., $\alphareheat$) and the
strength of disk feedback also play an important role. In particular, there
is significant shift in the median value of Var~15 towards smaller values
when higher redshift data is considered,
which implies, simultaneously, an increase in $\betadisk$ and a decrease in both
$\alphahot$ and $\nusf$. The combined effect is to reduce the efficiency of star formation in galaxy disks.
}

\subsection{The star formation history of the Universe}

In this paper we have deliberately focused on the GSMF.  This encoded the star
formation history of the Universe in the fossil record of the stars that have been
formed. It is nevertheless of interest to examine the star formation histories of
the models that have been selected on this basis.
Furthermore, it is interesting to separate models in which the mass loading in
starbursts, $\betaburst$, is comparable to that during quiescent star formation
($\betadisk$).
For simplicity, previous versions of \glf have assumed that the parameters for the
normalization of the mass loading in quiescent discs, $\betadisk$, and starbursts,
$\betaburst$,  were equal. By relaxing this assumption in this work, we found in
\S\ref{sec:subspace} that a larger $\betaburst$ is favoured.
While it is possible to find
plausible models for which
$\betaburst\sim \betadisk$, we found that most of the volume
(and the most plausible runs) of the plausible parameter space has
$\betaburst \gg \betadisk$.
\green{Since starbursts are more frequent at earlier times, it is worth noting that
a $\betaburst > \betadisk$ can lead to stronger supernova feedback at high redshift.
}

\begin{figure}
\includegraphics[width=\columnwidth]{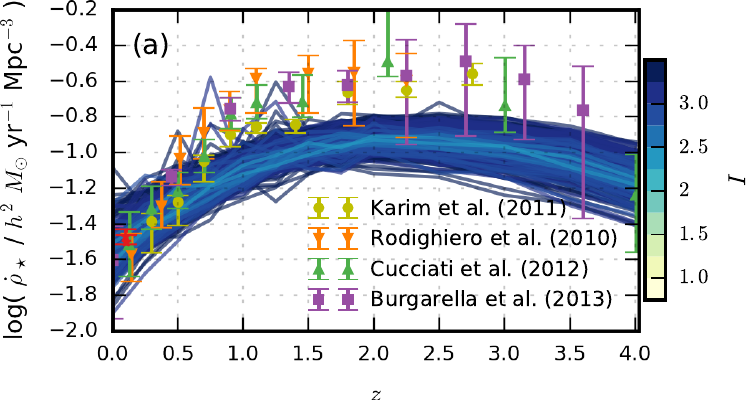}
\includegraphics[width=\columnwidth]{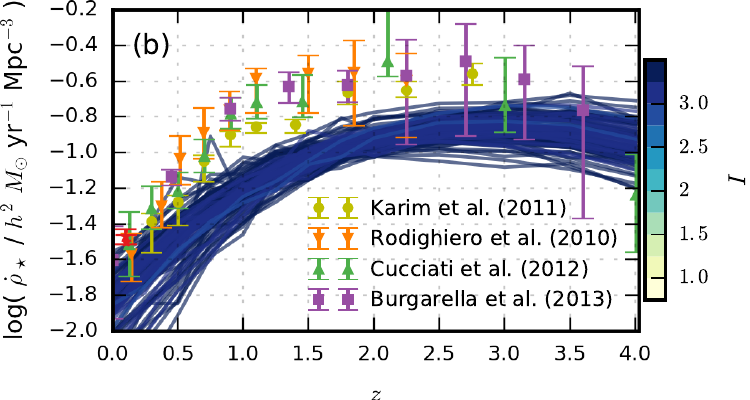}
\includegraphics[width=\columnwidth]{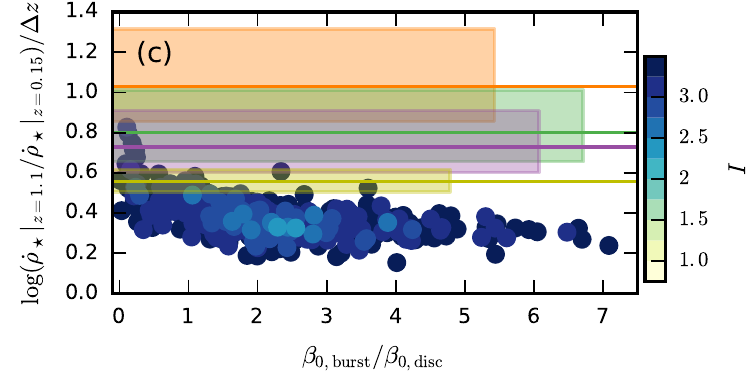}
\caption{Panels (a) and (b), show the cosmic star formation history (or
         comoving star formation rate density, SFRD), of
         runs with $I<3.5$ with respect to the redshifts
         $z=0.0,\,0.35,\,0.62,\,0.75,\,0.88,\,1.12$ and $1.25$. The
         colours correspond to their implausibilities as indicated.
         Observational data from
         \citet{Rodighiero2010,Karim2011,Cucciati2012}, and \citet{Burgarella2013}.
         Panel (a) shows only models with $\betaburst>2\,\betadisk$ while the
         panel (b) shows the case $\betaburst \leq \betadisk$.
         Panel (c) highlights the slope as a function of the $\betaburst/\betadisk$
         ratio with observational constraints shown as shaded areas (same colours as
         previous panels).
         While runs with a larger $\betaburst/\betadisk$ {display a}
         {qualitatively a better fit} to the
         SFRD, they fail to produce the strong increase in SFRD with redshift
         between $z=0.5$ and $z=1$, despite providing a better match to the GSMF
         evolution at the same redshift interval (as it can be seen by the colours).}
\label{fig:SFH}
\end{figure}

In Fig.~\ref{fig:SFH} we show the evolution of the cosmic star formation rate density
(SFRD) for runs with $\betaburst > 2\,\betadisk$  (upper panel) and for
$\betaburst \leq \betadisk$ (middle panel), in both cases selecting only
``acceptable'' runs,
with $I<3.5$ when conditioned on the full range of GSMF data. A selection of
observational data are shown as coloured points. Runs with a larger
$\betaburst/\betadisk$ ratio match well the observations for the SFRD at low
redshifts ($z\leq 0.5$),
but fail to reproduce the steep rise in SFRD with redshift in the interval
$0.5<z<1.0$.
When runs with $\betaburst \leq \betadisk$ are examined, one finds a stronger
redshift evolution of the SFRD, but the normalization is a factor $\sim 3$ off.  It
should be remembered, however, that the observational data does not measure the star
formation rate directly,
but requires calibration.  This is usually based on \cite{Kennicutt1998}. However, a
more recent study
by \cite{Chang2015} has suggested that this calibration needs revision (for Mid-IR
indicators),
bringing the observational data into better agreement with the $\betaburst \leq
\betadisk$ models.
A similar discrepancy was found in the numerical Eagle simulations
\green{\citep{Furlong2015} and other semi-analytic models (\citealt{Henriques2015},
see also \citealt{Guo2016}).}
It is noticeable, however, that these runs \green{-- Panel~(b) --} are generally in
less plausible agreement
with the mass function data than those in Panel~(a).

We quantify these differences more clearly in Panel~(c). Here the slope of the star
formation rate
rate density is plotted as a function of $\betaburst/\betadisk$, with the colour
coding indicating
the plausibility of the run. The coloured shaded regions indicate the constraints
implied by the observational data. The tension between the GSMF and the observed
decline in the star formation rate density is now evident. {While} models
require very small ratios {of $\betaburst/\betadisk$} to match the {SFRD}
observations{,
the most plausible models with respect to the GSMF evolution, i.e. those with
$I<2.5$, all have $1.66<\betaburst/\betadisk<2.56$.}

\section{Summary and Conclusions}
\label{sec:summary}

In this work, using an iterative emulator technique, we explored how the parameter
space of \glf is constrained by the galaxy stellar mass function.
After 6 waves of emulation, using only the local Universe GSMF data, more
than $99.9$ per cent of initial volume of the parameter space was deemed too
implausible for further exploration and was eliminated.
It was possible to find many parameter choices which provide a good match to
the local GSMF. The bi-variate projections of this space are shown
above the diagonal in Fig.~\ref{fig:implaus_space}.  The shape of the GSMF is
primarily  controlled by parameters related to star formation and feedback, namely:
$\alphacool$, $\alphahot$, $\betadisk$, $\alphareheat$, $\nusf\,$ {and}
$\,\fstab$.
Constraints on other parameters are weak.

We then included the requirement that the models also match higher
redshift data. This parameter space is shown below the diagonal of
Fig.~\ref{fig:implaus_space}.
This proves to be a much more stringent constraint, and the only acceptable runs
found had {$I\gtrsim 2$}.
The high mass end of the GSMF produced by this version of \glf is
typically too shallow for the local universe and becomes too steep at higher
redshifts.
This tension is a consequence of the observational data being consistent
with small or no increase in the abundance of high mass galaxies between
$z>0.35$ and $z=0.0$, compared to the model in which galaxies cannot avoid
growing in mass. This tension would still be present even if mass errors had been
underestimated by a factor of 2.

In order to better understand the dimensionality and most important variables of
the parameter space, we performing a PCA of the non-implausible volume of the
parameter space (constrained using the full range of redshifts,
Fig.~\ref{fig:pca_pairs}). We show that it is possible to write approximate
relations between the parameters, expressing conditions which need to be satisfied
in order to obtain a model with an acceptable match to the GSMF. Two principal
components (i.e. 2 directions in the parameter space) contain most of the
information about the basic shape of the GSMF, and these are mainly combinations of
the parameters $\alphacool$, $\alphahot$, $\betadisk$, $\alphareheat$, i.e. the
parameters controlling feedback processes.
The parameters  $\nusf$, $\fstab$ are also significantly constrained compared to
their initial values.

The PCA analysis provides a simple way to better understand why some model are able
to match both the local and high redshift GSMF data (points below the diagonal in
Fig.~\ref{fig:pca_pairs}),
while other models only match the observational at $z=0$ (points above the diagonal
in {Fig.~\ref{fig:pca_pairs}).
We show that the primary differences are encoded in Var~15, 16 and (primarily) 17.
Models which
match the $z=0$ GSMF but not the higher redshift data tend to have higher values of
Var~17, and thus lower values of $\fstab$ which tends to make disks more unstable at
low redshift.

{
In this paper, we explored a model in which the we allowed the mass loading in
starburst (driven by mergers or disk instabilities) to be different from the mass
loading in quiescent star formation.
The normalization of the quiescent mass, $\betadisk$ loading is strongly
constrained, while marginally acceptable models can be found for most of
the range of values for the burst mass loading, $\betaburst$. Nevertheless, this
does not mean that the full range of $\betaburst$ is equally plausible: there is a
much larger density of acceptable models ($I\lesssim 3$) with $20<\betaburst<30$
and the most plausible models, with $I<2.5$, have $1.66<\betaburst/\betadisk<2.56$.
}

We have deliberately focused the paper on the GSMF. This encoded the star formation
history of the
Universe, but we can also compare the models to the observed star formation rates of
galaxies. We do this by computing the volume averaged star formation rate density in
the model. We find that the star formation history is sensitive to the choice of the
ratio $\betaburst/\betadisk$. While models with $\betaburst>\betadisk$  offer a
reasonable match to the GSMF evolution,
they fail to display sufficiently rapid increase in the cosmic SFRD.
These results show the important additional information that can be extracted by
confronting the constrained models with additional datasets, but that this needs to
be done with care, since it is quite possible that systematic differences may make
it hard to simultaneously provide a plausible description of all the available data
if the observational uncertainites are taken at face value. The apparent
contradictions inherent in different datasets must be carefully accounted for: as
they may point to missing physics in the model. Clearly a future avenue for further
progress is to apply the methods we have developed here to a much wider range of
datasets.

Finally, we note that the main aim of this paper has been to examine how information
on the formation of galaxies can be extracted from observational dataset. We have
shown how simple physical results can emerge from the analysis of a highly
complex model. This approach can equally be applied across a wide range of science
disciplines where observational data are used to constrain seeming complex numerical
models.

\appendix

\section{Details of the History Matching procedure}
\label{ap:emu_details}
At each wave a set of
5000 runs were performed using space filling designs based on
maximin Latin hypercubes with rejection
\cite[see, for example,][]{SWMW89_DACE,Santner03_DACE,Currin91_BayesDACE}.
Third order polynomials were used as
the set of candidate regression terms for
$\beta_{ij} g_{ij}(x_{A_i})$ in equation~\eqref{eq_emulator}, with linear model
selection based on AIC criteria used to choose both the list of active inputs
$x_{A_i}$,
and the final list of polynomial terms used, for each output labelled by $i$.  As we
had access to reasonable numbers of runs at each wave we used a vague prior limit for
the $\beta_{ij}$ parameters and corresponding OLS estimates for the total residual
variance $\sigma_i^2 = \sigma_{u_i}^2 +  \sigma_{v_i}^2$, with $\sigma_{v_i}^2 =
\alpha \sigma_i^2$ where $\alpha$ was chosen so the nugget term represented a small
proportion of the total variance, checked using emulator diagnostics~\citep{Tony_EmDiag}. The correlation
lengths were specified to be $\theta_i = 0.35$, following the argument for the
residual of a third order polynomial fit presented by
\citet{Vernon2010}. The set of
outputs to be used in each wave $Q_k$ was chosen by scanning through all possible
outputs with approximate linear model regression based emulators, and selecting those that had the highest chance
of input space reduction, which were then emulated in detail using
equation~\eqref{eq_emulator}.

\section{Full PCA results}
\label{ap:fullPCA}
Table \ref{tab:pca} shows the full results of the PCA of the volume of the
parameter space containing models with $I<3.4$ with respect to redshifts
$z=0,\,0.35,\,0.62,\,0.75,\,0.88,\,1.12,\,1.25$ and $1.43$ (which corresponds to
508 runs in our library).
\begin{table*}
\begin{center}
\caption{
Principal component analysis for the acceptable space of galaxy stellar mass
functions (see details in the text). Each column shows one PCA variable, ordered {here} by
increasing standard deviation. Small relative standard deviations correspond to
components that are tightly constrained by the requirement
of producing a good luminosity function. Dominant input variables in each of the
vectors are highlighted in bold font. The variables have
been ordered so that the most constrained components appear {first}.
}
\begin{tabular}{cccccccccccccccccccccc}
\\ \hline 	& Mean & Var20 & Var19 & Var18 & Var17 & Var16 & Var15 & Var14 & Var13 & Var12 & Var11 \\  \hline
$F_{\rm SMBH}$ & 0.0357 & -0.00148 & 0.00467 & 0.0113 & -0.0102 & -0.0557 & 0.0538 & -0.0076 & -0.132 & \textbf{0.35} & 0.112 \\
$P_{\rm sf}$ & -0.0176 & -0.0116 & 0.0118 & -0.00198 & -0.0417 & 0.0764 & 0.0503 & 0.0777 & -0.293 & -0.274 & -0.134 \\
$V_{\rm cut}$ & 0.0341 & 0.00623 & 0.00514 & 0.0309 & -0.0103 & -0.07 & 0.00895 & -0.0931 & -0.121 & -0.0229 & 0.0929 \\
$\alpha_{\rm cool}$ & 0.0647 & \textbf{-0.634} & \textbf{-0.576} & -0.229 & 0.029 & \textbf{-0.322} & 0.226 & 0.0426 & -0.0754 & -0.0283 & -0.0785 \\
$\alpha_{\rm hot}$ & \textbf{0.673} & \textbf{0.583} & -0.232 & \textbf{-0.604} & -0.243 & -0.0284 & \textbf{0.404} & 0.03 & 0.0473 & 0.035 & -0.00876 \\
$\alpha_{\rm reheat}$ & \textbf{0.462} & 0.0444 & \textbf{0.356} & -0.0815 & -0.0894 & \textbf{-0.843} & -0.134 & 0.0222 & 0.0401 & 0.212 & 0.0795 \\
$\alpha_{\rm rp}$ & 0.14 & -0.0193 & -0.0123 & 0.00722 & 0.064 & 0.0411 & -0.0134 & 0.2 & \textbf{0.494} & -0.0779 & 0.0525 \\
$\beta_{0,{\rm burst}}$ & 0.0934 & -0.141 & -0.157 & 0.158 & -0.0282 & 0.188 & 0.164 & -0.0453 & 0.136 & \textbf{0.431} & \textbf{0.357} \\
$\beta_{0,{\rm disk}}$ & \textbf{-0.464} & \textbf{0.401} & \textbf{-0.669} & 0.26 & 0.079 & -0.172 & \textbf{-0.509} & -0.048 & -0.0144 & 0.0534 & 0.0391 \\
$\beta_{\rm sf}$ & -0.0574 & 0.00181 & 0.00683 & -0.00788 & 0.071 & -0.0104 & -0.124 & 0.222 & \textbf{0.364} & 0.24 & -0.29 \\
$\epsilon_{\rm edd}$ & -0.103 & 0.0123 & -0.00878 & 0.0584 & 0.0087 & -0.089 & -0.0217 & -0.242 & 0.175 & -0.0872 & -0.238 \\
$\epsilon_{\rm strip}$ & 0.0167 & 0.0143 & 0.00422 & -0.0143 & -0.0706 & -0.0291 & -0.0867 & \textbf{-0.315} & \textbf{-0.356} & 0.0492 & -0.133 \\
$\nu_{0,{\rm sf}}$ & \textbf{-0.456} & 0.239 & -0.0577 & \textbf{0.613} & 0.191 & -0.244 & \textbf{0.637} & 0.148 & -0.0399 & -0.0908 & -0.0714 \\
$\tau_{\rm min,burst}$ & -0.181 & -0.0143 & -0.00137 & 0.063 & -0.017 & 0.0348 & 0.191 & \textbf{-0.768} & 0.237 & 0.247 & -0.16 \\
$f_{\rm burst}$ & -0.106 & 0.00122 & 0.00577 & -0.0279 & 0.0264 & 0.096 & 0.017 & 0.107 & 0.148 & 0.23 & \textbf{0.36} \\
$f_{\rm dyn}$ & 0.062 & 0.00556 & 0.00795 & 0.00359 & 0.049 & 0.123 & -0.00325 & 0.216 & -0.164 & \textbf{0.432} & \textbf{-0.603} \\
$f_{\rm ellip}$ & -0.0319 & -0.00621 & -0.00332 & 0.0121 & -0.0198 & -0.0612 & -0.0399 & -0.206 & \textbf{0.307} & \textbf{-0.422} & -0.0321 \\
$f_{\rm stab}$ & \textbf{-0.362} & -0.0922 & -0.0586 & 0.297 & \textbf{-0.931} & 0.017 & -0.0108 & 0.106 & 0.0732 & -0.00902 & -0.0671 \\
$p_{\rm yield}$ & -0.175 & 0.0974 & 0.0909 & 0.102 & -0.0257 & 0.0788 & -0.0834 & -0.0576 & -0.222 & 0.0669 & 0.0311 \\
$z_{\rm cut}$ & -0.0931 & -0.00263 & -0.000888 & 0.0131 & -0.0299 & -0.00443 & 0.0278 & -0.0683 & -0.255 & 0.0149 & \textbf{0.357} \\
\hline Rel. Std. Dev.
&	& 0.0707 & 0.0952 & 0.174 & 0.217 & 0.354 & 0.361 & 0.405 & 0.442 & 0.461 & 0.469 \\
\hline\hline
	& Var10 & Var9 & Var8 & Var7 & Var6 & Var5 & Var4 & Var3 & Var2 & Var1 \\ \hline
$F_{\rm SMBH}$ & -0.127 & \textbf{0.368} & \textbf{0.431} & 0.279 & -0.175 & -0.111 & \textbf{0.349} & \textbf{-0.306} & -0.15 & \textbf{-0.381} \\
$P_{\rm sf}$ & \textbf{0.37} & \textbf{-0.422} & \textbf{0.353} & 0.262 & \textbf{0.39} & 0.00731 & -0.0154 & \textbf{-0.362} & -0.111 & 0.00302 \\
$V_{\rm cut}$ & \textbf{0.534} & 0.0675 & -0.144 & 0.0874 & \textbf{-0.374} & -0.0474 & -0.12 & 0.241 & \textbf{-0.655} & -0.0106 \\
$\alpha_{\rm cool}$ & -0.0962 & -0.0286 & -0.113 & 0.0446 & -0.0366 & -0.0254 & 0.0225 & -0.0839 & -0.0507 & 0.0843 \\
$\alpha_{\rm hot}$ & -0.0202 & -0.0176 & 0.0152 & 0.0385 & -0.0485 & 0.0562 & -0.0579 & -0.00184 & -0.0117 & 0.0512 \\
$\alpha_{\rm reheat}$ & 0.0525 & -0.186 & 0.00651 & -0.037 & 0.0899 & 0.134 & -0.0759 & -0.0512 & 0.0341 & -0.0262 \\
$\alpha_{\rm rp}$ & -0.2 & 0.00332 & 0.215 & \textbf{-0.358} & 0.164 & 0.233 & 0.0666 & -0.22 & \textbf{-0.578} & 0.132 \\
$\beta_{0,{\rm burst}}$ & 0.0889 & -0.246 & 0.265 & 0.129 & -0.0482 & \textbf{0.436} & \textbf{-0.381} & 0.147 & 0.131 & -0.0425 \\
$\beta_{0,{\rm disk}}$ & 0.0519 & -0.0732 & 0.00204 & -0.0501 & 0.0594 & -0.0166 & 0.0345 & -0.0425 & 0.00823 & -0.0667 \\
$\beta_{\rm sf}$ & 0.0626 & 0.136 & 0.132 & \textbf{0.349} & -0.0256 & \textbf{-0.456} & \textbf{-0.413} & -0.0844 & 0.019 & \textbf{0.328} \\
$\epsilon_{\rm edd}$ & 0.161 & 0.0676 & 0.258 & 0.22 & -0.206 & \textbf{0.35} & \textbf{0.469} & 0.157 & 0.15 & \textbf{0.507} \\
$\epsilon_{\rm strip}$ & -0.27 & \textbf{0.425} & 0.0524 & 0.125 & \textbf{0.449} & 0.231 & \textbf{-0.307} & 0.166 & -0.237 & 0.186 \\
$\nu_{0,{\rm sf}}$ & -0.0611 & 0.0929 & -0.067 & 0.0272 & 0.059 & -0.0342 & -0.0325 & 0.0103 & -0.00214 & 0.0434 \\
$\tau_{\rm min,burst}$ & 0.0595 & -0.211 & -0.0785 & -0.171 & 0.136 & -0.283 & 0.0462 & -0.18 & -0.0739 & -0.0642 \\
$f_{\rm burst}$ & 0.221 & 0.141 & \textbf{-0.523} & \textbf{0.319} & \textbf{0.422} & 0.0604 & \textbf{0.314} & -0.152 & -0.0191 & 0.161 \\
$f_{\rm dyn}$ & 0.23 & 0.0689 & -0.218 & -0.262 & -0.0617 & \textbf{0.351} & 0.00458 & -0.22 & 0.0228 & -0.0977 \\
$f_{\rm ellip}$ & 0.0984 & 0.289 & -0.171 & 0.225 & -0.157 & \textbf{0.322} & \textbf{-0.305} & \textbf{-0.413} & 0.142 & \textbf{-0.304} \\
$f_{\rm stab}$ & -0.000828 & 0.0418 & -0.043 & -0.0288 & 0.00779 & -0.0626 & 0.0502 & 0.000526 & -0.0122 & 0.00991 \\
$p_{\rm yield}$ & \textbf{-0.473} & \textbf{-0.371} & -0.274 & 0.27 & \textbf{-0.375} & 0.0953 & -0.0405 & \textbf{-0.348} & -0.198 & 0.281 \\
$z_{\rm cut}$ & 0.229 & 0.275 & 0.122 & \textbf{-0.429} & -0.14 & -0.0977 & -0.141 & \textbf{-0.425} & 0.189 & \textbf{0.456} \\
\hline Rel. Std. Dev.
& 0.476 & 0.497 & 0.507 & 0.522 & 0.533 & 0.542 & 0.566 & 0.582 & 0.596 & 0.619 \\
\hline

\end{tabular}
\label{tab:pca}
\end{center}
\end{table*}

\section*{Acknowledgements}
We thank Cedric Lacey for comments on the paper.
LFSR has been supported by STFC (ST/N000900/1 and ST/L005549/1) and acknowledges
support from the European Commission's Framework Programme 7, through the Marie Curie
International Research Staff Exchange Scheme LACEGAL (PIRSES-GA-2010-269264).
LFSR thanks Jacqueline Dourado and the Federal University of Piau\'i for the
kindness and hospitality during the writing of part of this work.
IV gratefully acknowledges MRC (RF060151) and EPSRC (EP/E00931X/1) funding.
The research was also supported by the UK Science and Technology Facilities Council
(ST/F001166/1 and ST/I000976/1), Rolling and Consolidating Grants to the ICC.
This work used the DiRAC Data Centric system
at Durham University, operated by the Institute for Computational Cosmology on
behalf of the STFC DiRAC HPC Facility (www.dirac.ac.uk). This equipment was funded by
BIS National E-infrastructure capital grant (ST/K00042X/1), STFC capital grant
(ST/H008519/1), and STFC DiRAC Operations grant (ST/K003267/1) and Durham University.
DiRAC is part of the National E-Infrastructure.
This research has made use of NASA's Astrophysics Data System.

\footnotesize{
    \bibliographystyle{mnras}
    \bibliography{glf_explorer}
}
\bsp
\label{lastpage}
\end{document}